\documentclass[sigconf, anonymous]{acmart}

\usepackage{amsmath,amsfonts}
\usepackage{graphicx}
\usepackage{textcomp}
\usepackage{xcolor}
\usepackage{epsfig,subcaption}
\usepackage{tikz}
\usepackage{bm}
\usepackage{enumitem}
\setenumerate[1]{itemsep=0pt,partopsep=0pt,parsep=\parskip,topsep=5pt,leftmargin=10pt}
\setitemize[1]{itemsep=0pt,partopsep=0pt,parsep=\parskip,topsep=5pt,leftmargin=10pt}
\setdescription{itemsep=0pt,partopsep=0pt,parsep=\parskip,topsep=5pt,leftmargin=10pt}
\usepackage{algorithm}
\usepackage{algpseudocode}
\usepackage{diagbox}
\usepackage{makecell}
\usepackage{booktabs}
\usepackage{adjustbox}
\usepackage{latexsym,fancyhdr,hyperref}
\usepackage{natbib}
\usepackage{adjustbox,multirow, booktabs, hhline}

\def\hlb{\color{black}}
\newcommand{\removecontent}[1]{}

\def\TableCaptionSpacing{-0.3cm}

\def\BibTeX{{\rm B\kern-.05em{\sc i\kern-.025em b}\kern-.08em
    T\kern-.1667em\lower.7ex\hbox{E}\kern-.125emX}}
\begin{document}
\title{Perception-Aware Attack: Creating Adversarial Music via Reverse-Engineering Human Perception}

\author{Rui Duan}
\affiliation{%
    \institution{University of South Florida}
    \city{Tampa}
    \state{FL}
    \country{USA}
    }
\email{ruiduan@usf.edu}

\author{Zhe Qu}
\affiliation{%
    \institution{University of South Florida}
    \city{Tampa}
    \state{FL}
    \country{USA}
    }
\email{zhequ@usf.edu}

\author{Shangqing Zhao}
\affiliation{%
    \institution{University of Oklahoma}
    \city{Tulsa}
    \state{OK}
    \country{USA}
    }
\email{shangqing@ou.edu}

\author{Leah Ding}
\affiliation{%
    \institution{American University}
    \city{Washington}
    \state{DC}
    \country{USA}
    }
\email{ding@american.edu}

\author{Yao Liu}
\affiliation{%
    \institution{University of South Florida}
    \city{Tampa}
    \state{FL}
    \country{USA}
    }
\email{yliu@cse.usf.edu}

\author{Zhuo Lu}
\affiliation{%
    \institution{University of South Florida}
    \city{Tampa}
    \state{FL}
    \country{USA}
    }
\email{zhuolu@usf.edu}

\begin{abstract}
Recently, adversarial machine learning attacks have posed serious security threats against practical audio signal classification systems, including speech recognition, speaker recognition, and music copyright detection. Previous studies have mainly focused on ensuring the effectiveness of attacking an audio signal classifier via creating a small noise-like perturbation on the original signal. It is still unclear if an attacker is able to create audio signal perturbations that can be well perceived by human beings in addition to its attack effectiveness. This is particularly important for music signals as they are carefully crafted with human-enjoyable audio characteristics.

In this work, we formulate the adversarial attack against music signals as a new perception-aware attack framework, which integrates human study into adversarial attack design. Specifically, we conduct a human study to quantify the human perception with respect to a change of a music signal. We invite human participants to rate their perceived deviation based on pairs of original and perturbed music signals, and reverse-engineer the human perception process by regression analysis to predict the human-perceived deviation given a perturbed signal. The perception-aware attack is then formulated as an optimization problem that finds an optimal perturbation signal to minimize the prediction of perceived deviation from the regressed human perception model. We use the perception-aware framework to design a realistic adversarial music attack against YouTube's copyright detector. Experiments show that the perception-aware attack produces adversarial music with significantly better perceptual quality than prior work.

\end{abstract}

%


\maketitle

\section{Introduction}\label{Sec:Introduction}

Adversarial machine learning attacks, originated from the image domain \cite{goodfellow2014explaining,kurakin2016adversarial,carlini2017towards,szegedy2013intriguing}, have recently become a serious security issue in audio signal processing system designs leveraging machine learning, including speech recognition \cite{carlini2016hidden, yuan2018commandersong, chen2020devil, qin2019imperceptible, schonherr2018adversarial}, speaker identification \cite{chen2019real, abdullah2019hear}, and music copyright detection \cite{saadatpanah2020adversarial}.

Adversarial machine learning attacks attempt to create a small perturbation on the original audio signal such that a machine learning classifier can yield an incorrect output. For example, a small change in a speech command could make Amazon Echo \cite{test_amazon} and Google assistant \cite{test_Google} recognize a different, yet malicious command \cite{chen2020devil,zheng2021black}. And manipulating copyrighted music might bypass the copyright detection in YouTube \cite{saadatpanah2020adversarial}. One key component in adversarial audio signals is the perturbation, which is designed to cause misclassification and at the same time be small enough to be hardly noticed. To quantify the perturbation, existing studies \cite{chen2019real,li2020advpulse} usually use a mathematical distance (e.g., the Euclidean distance \cite{saadatpanah2020adversarial}, or more generally, the $L_p$ norm \cite{carlini2017towards}) between the original and perturbed audio signals. As a result, the perturbed signal with the minimized distance to the original one could be considered as a good candidate under the constraint that it can successfully spoof the classifier.

However, the $L_p$ norm based methods only measure the magnitude distance between two signals; but the human perception is much more complex than computing the magnitude distance. There exists a gap between the mathematical distance and the eventual human perception. Although the two may be related in some way (e.g., zero distance meaning no signal perturbation), there is still no direct relation to indicate an increase or decrease of the distance in mathematics would be human-perceived as the same. For example, adding a perturbation that is the same as the original music signal is equivalent to increasing the volume of the music, which does not quite change the human perception of music quality. Indeed, a few studies \cite{carlini2017towards,qin2019imperceptible} have pointed out similar issues and indicated that new methods are needed to measure the perceptual similarity between the original and perturbed signals; but there is limited work on systematically designing adversarial machine learning from the human perception perspective.

In this paper, we create a new mechanism to craft adversarial audio signals. We focus on generating adversarial music signals to bypass a music copyright detector and hardly raise human attention. To this end, we formulate the relationship between signal perturbation and human perception with two key steps: i) quantifying the change of human perception with respect to the change of a music signal; and ii) finding a new way to generate perturbations to minimize the change in human perception and fool a classifier.

To study how a change of a music signal affects human perception, we first conduct a human study where volunteers quantify their perceived deviations between the original and perturbed signals as ratings on a Likert scale \cite{thiede2000peaq}. We use regression analysis to build an approximate mathematical relation between the change of music and the human-perceived deviation rating obtained from the human study. Given a perturbed signal, we use the regressed model to predict the human rating on the perceived deviation. We call this output quantified deviation (qDev).



We then reformulate adversarial machine learning for music signals as a perception-aware attack problem of finding a perturbation that minimizes its qDev while misleading a target classifier. The reformulation, however, leads to a computationally intractable optimization with a non-convex and non-differentiable objective function. To solve this problem, we propose a method by reducing the search space for finding a feasible solution. We observe that a common process in music classification is to identify and extract audio fingerprints (e.g., high energy values on certain frequencies) from a signal's spectrogram \cite{wang2003industrial, cano2002robust, pardo2006finding}. Creating a perturbation may introduce additional frequencies and energy values, which will generate new fingerprints different from the original signal. Such difference can be used to fool the target classifier. Meanwhile, to make the perturbation less noticeable to humans, our proposed perception-aware attack is designed to create new frequencies and energy values as a perturbation to minimize the qDev metric. We show that the perception-aware attack can produce adversarial music more effectively in terms of attack success rate and human-perceived quality. We test our perception-aware attack on different genres of music against YouTube's copyright detection. 
Experimental results show that the perception-aware attack can produce effective adversarial music to bypass YouTube's detection while achieving a significantly higher perceptual quality compared to a recent $L_p$ norm based attack \cite{saadatpanah2020adversarial}.

Our major contributions are summarized as follows.
\begin{itemize}
    \item We conduct a human study to understand how human participants perceive the music signal perturbation. We use regression analysis to model the relationship between the audio feature deviation and the human-perceived deviation for music signals.
    \item Based on the regressed human perception model, we propose, formulate, and evaluate the perception-aware attack framework to create adversarial music.
    \item The perception-aware attack is able to perturb music signals with better perceptual quality and achieve higher attack success rates than conventional $L_p$ norm based attacks against YouTube's copyright detector.
    \item To the best of our knowledge, our study presents the first systematic work that integrates human factors into the internals of adversarial audio attacks. We believe the results will encourage further human-in-the-loop research.
\end{itemize}
The rest of the paper is organized as follows: Section~\ref{Sec:background} introduces the background and the motivation of our study. Section~\ref{Sec:Human Perception} elaborates our human study with regression analysis. We formulate the perception-aware attack framework, create a realistic attack, and conduct experiments in Sections~\ref{Sec:Attack}, \ref{Sec:RealAttack}, and \ref{Sec:Experiments}, respectively. Potential defense strategies are discussed in Section~\ref{Sec:Defense}. Finally, we summarize related work in Section~\ref{Sec:RelatedWork} and conclude this paper in Section~\ref{Sec:Conclusion}.

\section{Background and Design Motivation}\label{Sec:background}
In this section, we briefly introduce the background and describe our motivation and design intuition.

\subsection{Representation of Music Signal}\label{music_perception}
\begin{figure}[!t]
    \centering
    \includegraphics[width=0.46\textwidth]{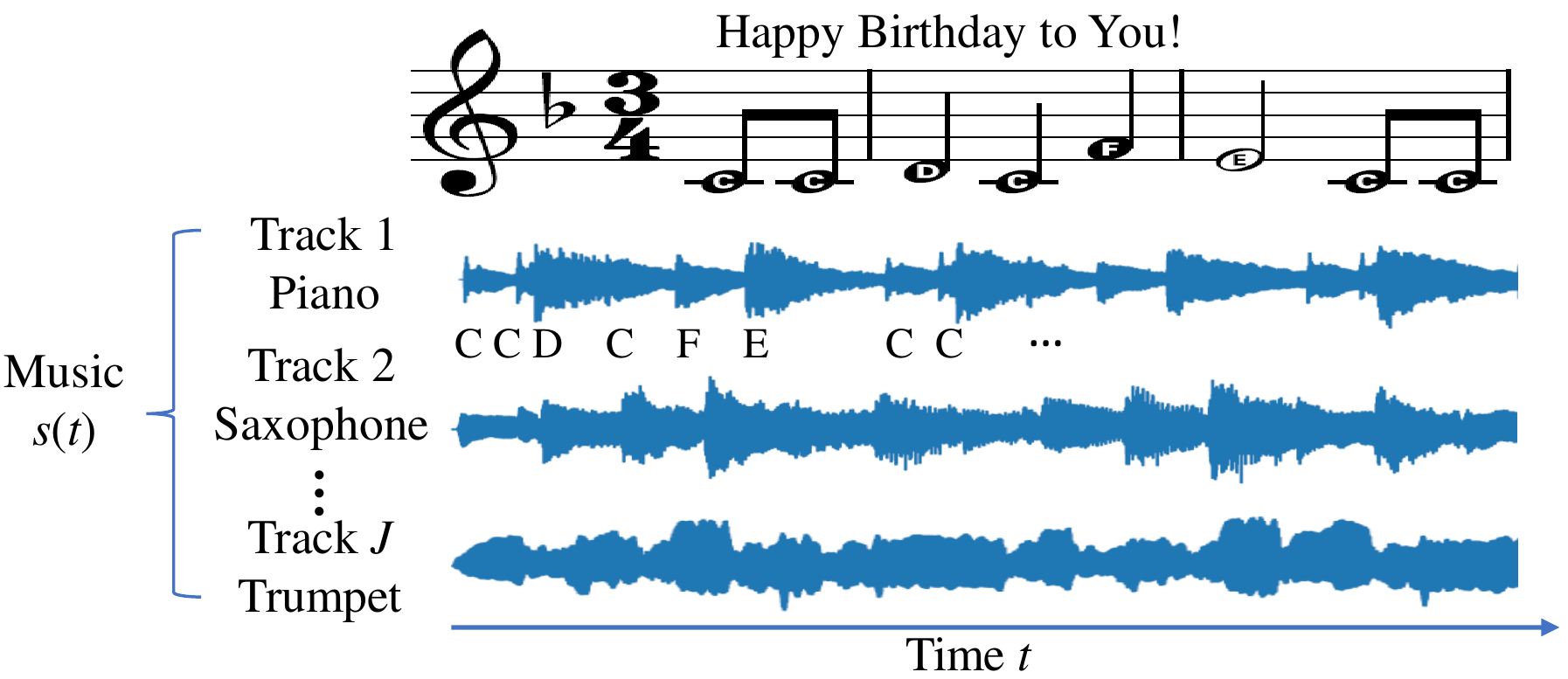}
    \caption{Music with multiple track signals by different instruments, and each track contains a series of notes.}
    \vspace{-0.0cm}
    \label{Fig:Notes}
\end{figure}

As an example shown in Fig.~\ref{Fig:Notes}, a digital music signal $s(t)$ at sample time $t \in \{0, 1, 2, \cdots, T\}$ (where $T$ is the number of signal samples) can be represented as the sum of audio track signals \cite{uhlich2017improving}, i.e., $s(t) = \sum_{j=1}^J s_j(t)$, where $J$ is the number of tracks, and the track signal $s_j(t)$ is a time-series of harmonic notes \cite{moorer1977signal,muller2011signal,kereliuk2007indirect,risset1999exploration}. A note, similar to a phoneme of speech \cite{zhang2017dolphinattack, yuan2018commandersong}, is the smallest signal unit of a piece of music consisting of a fundamental frequency and a set of harmonics \cite{godsill2002bayesian,godsill2003bayesian,walmsley1998multidimensional}. 

\subsection{Adversarial Audio Attacks}
Given a classifier with prediction function $f(\cdot)$ which takes the input audio signal $s(t)$ and outputs the correct label $ f(s(t))= y$, existing adversarial audio attacks \cite{carlini2018audio,yakura2018robust,qin2019imperceptible} aim to add a small signal perturbation $\delta(t)$ to the original audio signal $s(t)$, and then supply the perturbed signal $\hat s(t) \!=\! s(t) \!+\! \delta(t)$ to the classifier that accordingly generates an incorrect label. The method of creating $\delta(t)$, which mainly inherits from the fundamental framework in the image domain \cite{carlini2017towards,szegedy2013intriguing}, can be formulated as
\begin{eqnarray}
 \text{minimize}  &&   \| \delta(t) \|_p \label{Eq:TAML}\\
 \text{subject to} &&   f(\hat s(t))\neq y \notag,
\end{eqnarray}
where $\| \delta(t) \|_p$ denotes the $L_p$ norm of the perturbation $\delta(t)$ \cite{carlini2017towards,goodfellow2014explaining}. The objective of \eqref{Eq:TAML} is to minimize the change of the perturbed signal $\hat s(t)$ from the original $s(t)$. Since it is computationally difficult to solve \eqref{Eq:TAML}, many variants of formulating the adversarial audio attacks have been proposed for distinct attack scenarios, such as speech recognition \cite{carlini2018audio,yakura2018robust,qin2019imperceptible}, speaker recognition \cite{chen2019real,zheng2021black}, and music copyright detection \cite{saadatpanah2020adversarial}. To still make $\hat s(t)$ look like $s(t)$, these formulations limit the $L_p$ norm of the perturbation $\delta(t)$ within a given threshold $\epsilon$, i.e., $\| \delta(t) \|_p \leq \epsilon$.
%
%
The $L_{\infty}$, $L_{2}$, and $L_0$ norms are commonly adopted in the literature to create adversarial attacks targeting various audio signal classifiers \cite{carlini2018audio,li2020advpulse,zheng2021black, chen2019real,kreuk2018fooling}.

\subsection{Motivation and Design Intuition}
Although existing adversarial audio attacks mathematically limit the magnitude of the perturbation $\delta(t)$ via $\| \delta(t) \|_p \leq \epsilon$, it is still not clear whether such a constraint is the most effective to make the perturbation unnoticeable by human beings. For example, a few studies \cite{carlini2017towards,qin2019imperceptible} have noted the concern on whether the $L_p$ norm metric is appropriate to measure the signal similarity from the human perception perspective. In other words, there is no evidence to show that the deviation in human cognition can be represented by $\|\delta(t)\|_p$. As a result, we are motivated to investigate the problem. Our goals are twofold: i) relating the change of a music signal to the deviation of human perception and ii) finding a new way to create the perturbation that is unnoticeable by human beings as much as possible. To achieve these goals, our design consists of three major components.
\begin{enumerate}
\item {\it Reverse-engineering human perception of signal deviation,} we treat human perception as a black box and design a human study to quantify human perceived deviations. Specifically, we invite volunteers to assign a rating of perceived deviation to measure the difference between the original and perturbed signals. Then, we reverse-engineer the black box via regression analysis to build a relationship between the signal deviation and the human-perceived deviation.

\item {\it Reformulating the adversarial audio attack as the perception-aware attack,} based on the relationship found in the human study, we establish the perception-aware attack framework with the objective to quantitatively minimize the perceived deviation while attacking audio classification.

\item {\it Demonstrating a realistic attack against a music copyright detector,} based on the new attack framework, we create adversarial music against YouTube's copyright detector. We demonstrate via experiments the effectiveness of the attack in terms of success rate and human-perceived deviation.
\end{enumerate}


\subsection{Threat Model}
In this paper, we consider an attacker that aims to find a perturbation $\delta(t)$ to a music signal $s(t)$ such that $\hat s(t) \!=\! s(t) \!+\! \delta(t)$ leads to an incorrect output of an audio signal classifier, which is similar to the goal of existing audio attacks \cite{szegedy2013intriguing, carlini2018audio,yakura2018robust,li2020advpulse,zheng2021black, saadatpanah2020adversarial}. At the same time, the attacker is designed to be aware of how $\hat s(t)$ affects the human perception and minimizes its perceived deviation from $s(t)$. We assume that the attacker has no knowledge of the algorithm design or parameter choices in the classifier, but has access to the classification result of any input signal. We also assume that the attacker has no access to the classifier's training database. A representative commercial scenario is that an attacker wants to bypass YouTube's copyright detector \cite{saadatpanah2020adversarial} and use copyrighted music content in an unauthorized way to attract more online views for advertisement revenue gain.

\section{Reverse-Engineering Human Perception of Music Signals}\label{Sec:Human Perception}
In this section, we present how to quantify the human perceived deviation of music signals. We first analyze the key features for the signal quality, then conduct the human study, and lastly present the study results and regression analysis.

\subsection{Audio Features for Human Perception}\label{SubSec:Features}

Based on existing studies in audio engineering \cite{platel1997structural, tsai2011automatic, molina2013fundamental, gupta2017perceptual, law2012assessing}, there are four widely-used features: pitch, rhythm, timbre, and loudness. Pitch is the subjective perception of highness or lowness of a sound, and is referred to as the fundamental frequency $\omega_0$ of a note \cite{hartmann2004signals,loeffler2006instrument}. Rhythm is described as the tempo of the musical sound \cite{tsai2011automatic}, which depends on the length of each note and the time intervals between adjacent notes. Timbre is the mixture of the harmonics, which brings the ''color'' to music \cite{loeffler2006instrument,wessel1979timbre}, and it is similar to the characteristics of the speech \cite{de2012enhancing}. Loudness measures the intensity of an audio signal and can be seen as the energy level or the volume of the signal \cite{tsai2011automatic}.

\begin{figure}[!t]
  \centering
  \includegraphics[width=5.2cm,height=4.5cm]{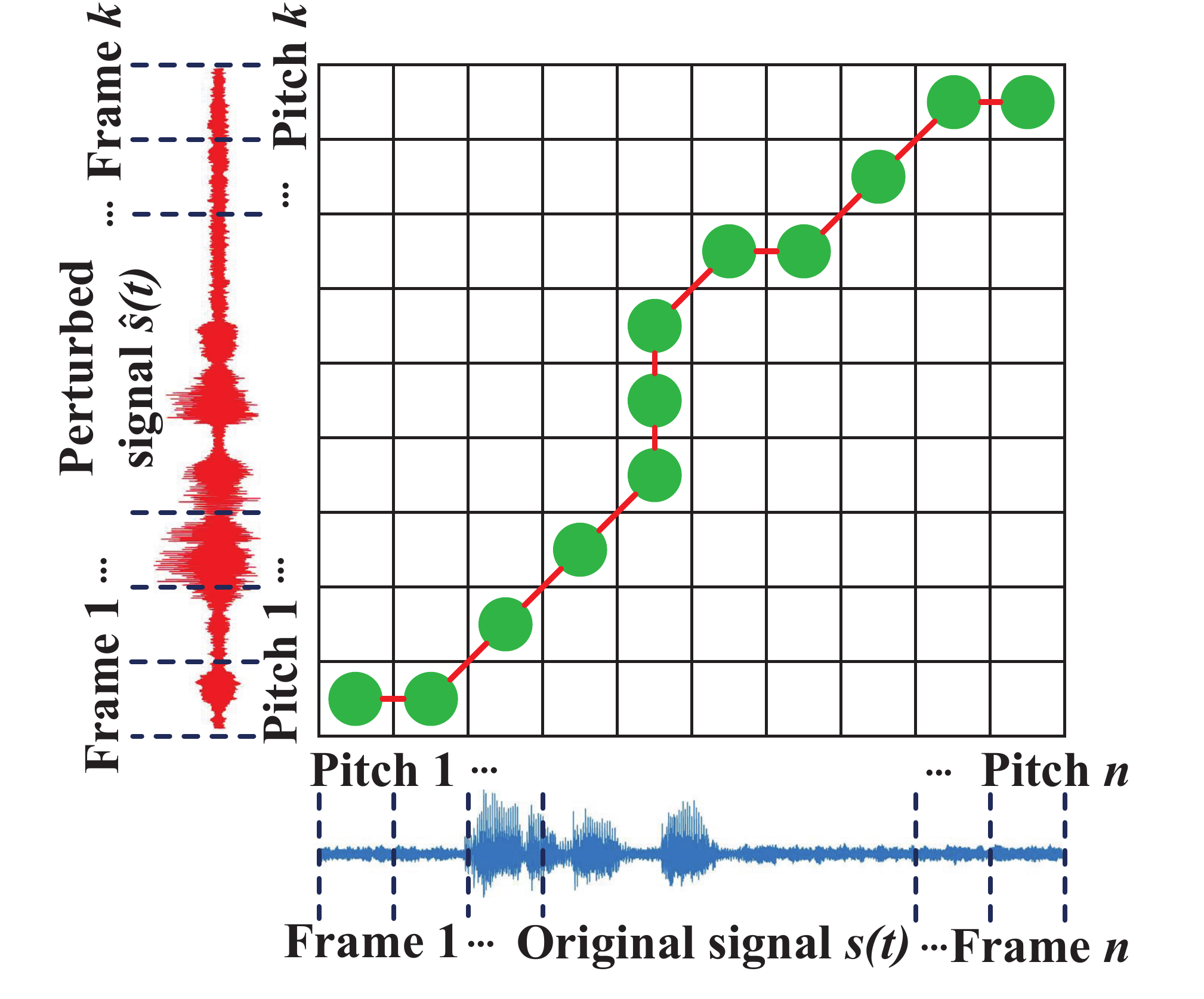}
  \caption{Computing deviation values via DTW.}
  \label{Fig:DTW}
\end{figure}

In the following, we briefly introduce the commonly-used methods to compute the feature deviations between two signals $s(t)$ and $\hat s(t)$ in the literature. For each feature, the procedure is the same and shown in Fig.~\ref{Fig:DTW}: $s(t)$ and $\hat s(t)$ each will be separated into frames with a small time interval (e.g., 16ms \cite{gupta2017perceptual}). The signal samples in each frame are used to generate a feature value (e.g., pitch value). The feature values from all frames constitute a time-series data vector. Then, an algorithm called Dynamic Timing Warping (DTW) \cite{sakoe1978dynamic,salvador2007toward} is used to quantify the similarity between the time-series vector for $s(t)$ and the one for $\hat s(t)$, and generate a vector of frame-wise deviation values for the feature. The advantage of DTW over the Euclidean distance is that DTW can reduce the time distortion \cite{ratanamahatana2004making} via finding an optimal path between two time-series vectors. For instance, the red line in Fig.~\ref{Fig:DTW} indicates the DTW path between $s(t)$ and $\hat s(t)$.

\begin{figure*}[!t]
\centering
\begin{subfigure}{0.45\columnwidth}
\includegraphics[width=\columnwidth]{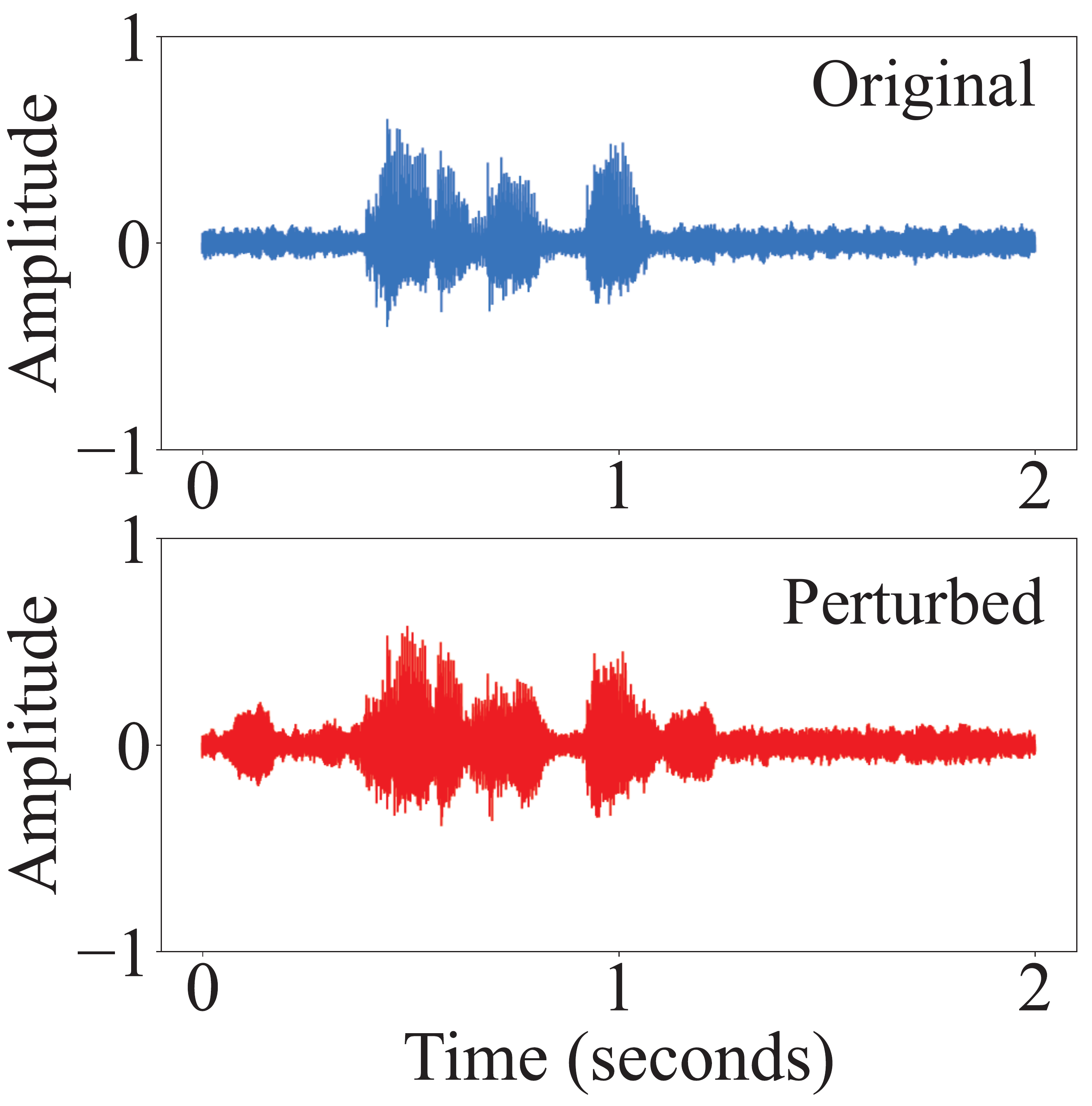}
\caption{Waveforms}\label{fig:waveform}
\end{subfigure}%
\hfill
\begin{subfigure}{0.455\columnwidth}
\includegraphics[width=\columnwidth]{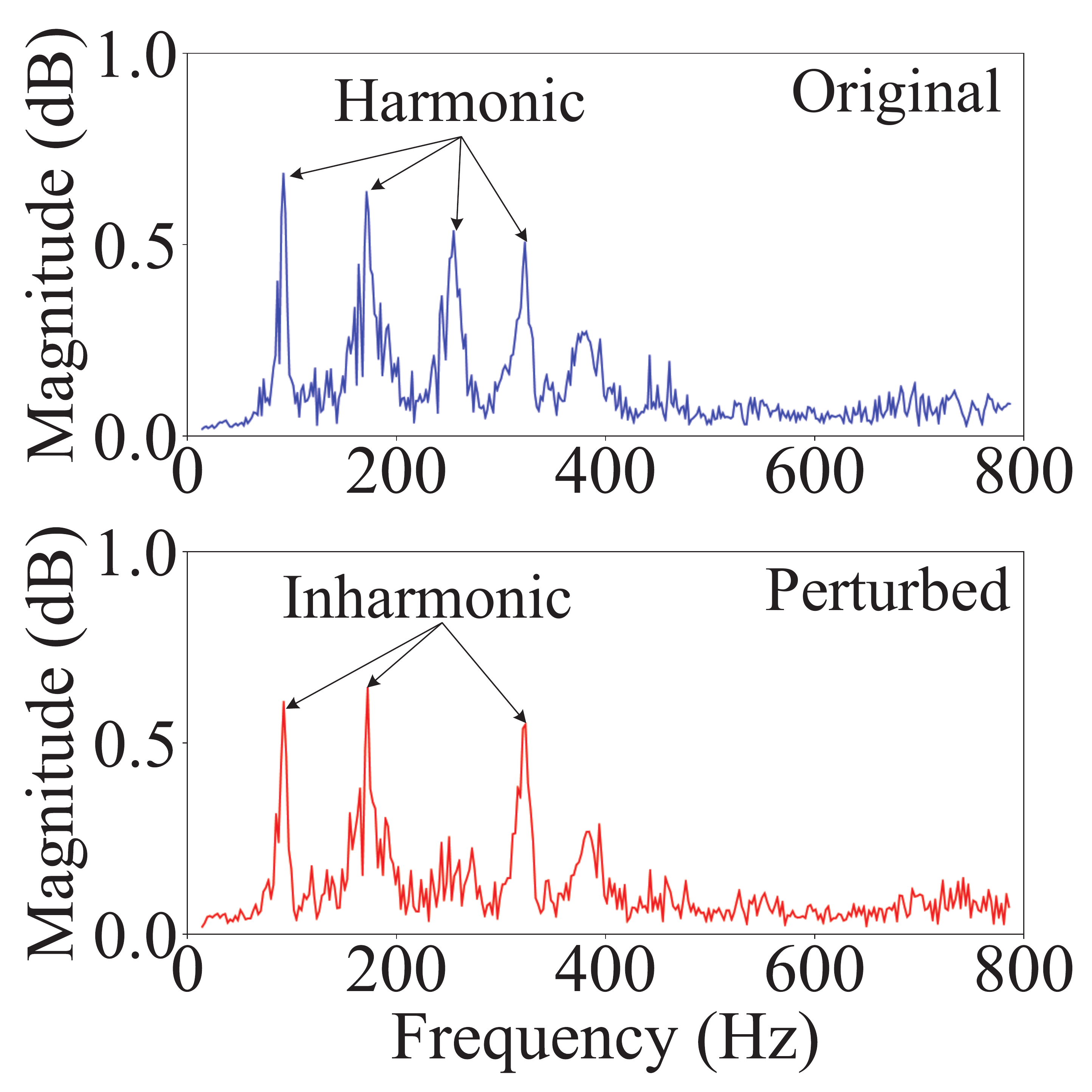}
\caption{Harmonic note spectrum}\label{fig:Original_music_spectrum}
\end{subfigure}%
\hfill
\begin{subfigure}{0.5\columnwidth}
\includegraphics[width=\columnwidth]{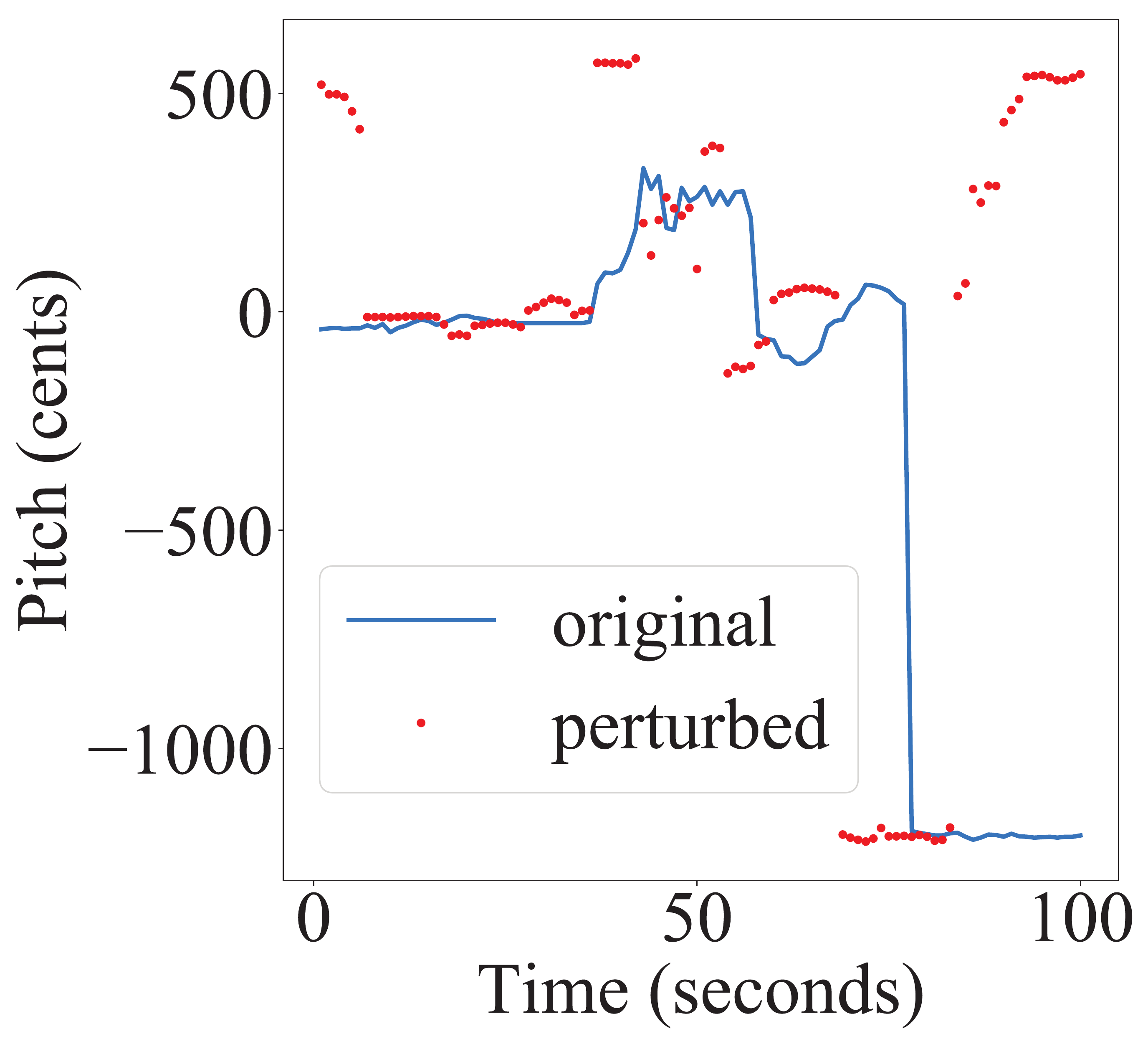}
\caption{Pitch deviation}\label{fig:Pitch_deviation}
\end{subfigure}%
\hfill
\begin{subfigure}{0.5\columnwidth}
\includegraphics[width=\columnwidth]{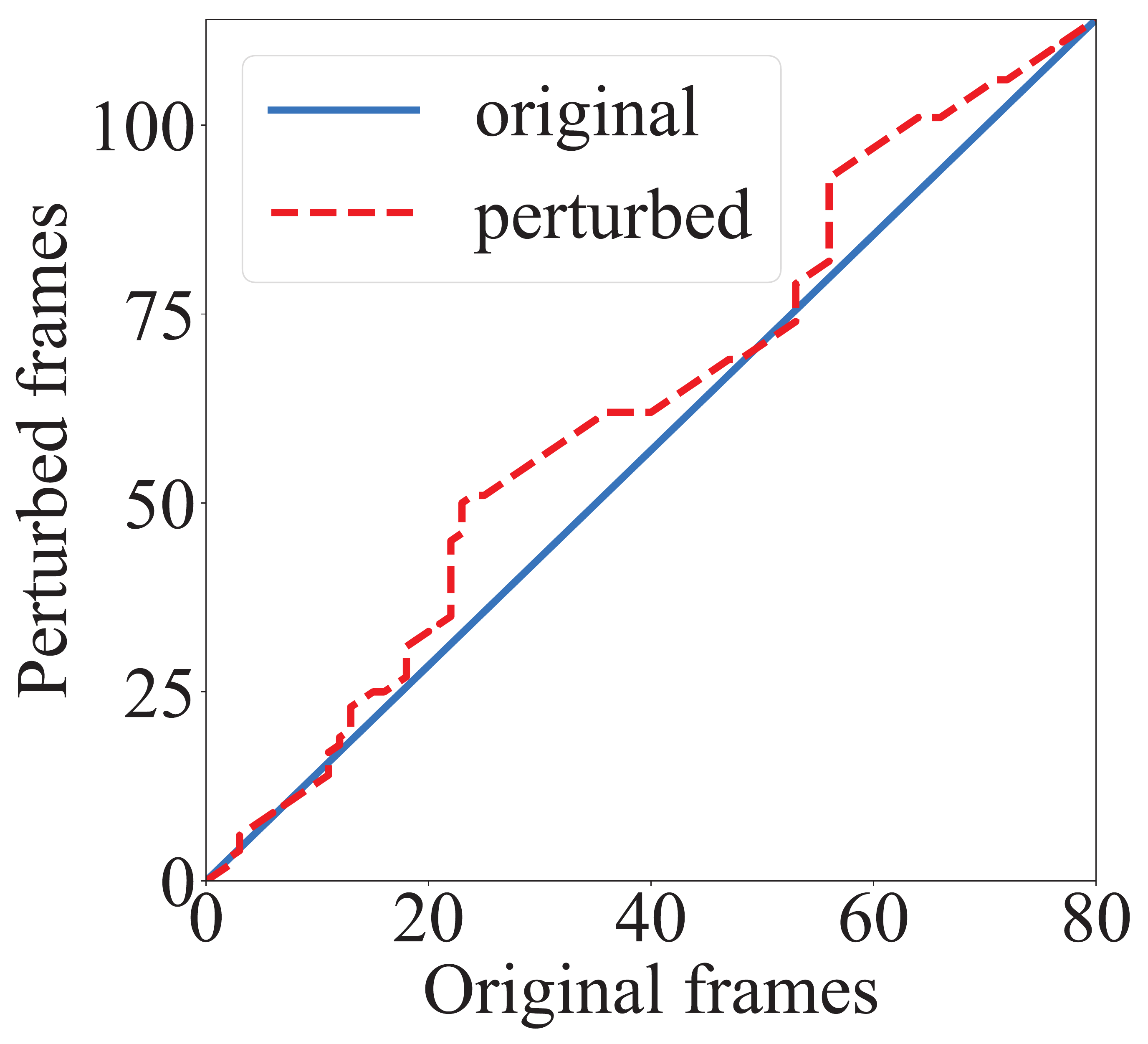}
\caption{Rhythm deviation}\label{fig:Rhythm_deviation}
\end{subfigure}%
\vspace{-0.2cm}
\caption{Impacts of a noise-like perturbation on the music features: a 2-second attack example ``Boom Boom Pow'' from existing work \cite{saadatpanah2020adversarial}. Specifically, (a) and (b) shows the waveforms and spectrums, respectively; (c) and (d) show the pitch contours and the rhythm DTW paths between the perturbed and original signals, respectively.}
\vspace{-0.1cm}
\label{fig:key_insights}
\end{figure*}

\begin{itemize}
\item Pitch: The pitch value in each frame is the basic frequency $\omega_0$ obtained via pitch estimation, which is a maximum likelihood estimation problem \cite{duan2010multiple} via finding $\omega_0$ from harmonics $\sum^M_{m=1}m\omega_0$. The estimated pitch values from all frames form a time series for each signal and then DTW is used to generate the vector of frame-wise pitch deviation values between the two signals. 

\item Rhythm: Rhythm computation is based on pitch estimation. A deviation value for rhythm between two frames is computed as the linear regression error in DTW during computing the deviation value for pitch \cite{molina2013fundamental}. All these values generated during DTW form the vector of frame-wise deviation values for rhythm.

\item Timbre: The timbre value for each frame is computed as a Mel-Frequency Cepstrum Coefficient (MFCC) \cite{davis1980comparison}. The vector of frame-wise deviation values for timbre is the result of the DTW between the MFCC vectors for $s(t)$ and $\hat s(t)$.


\item Loudness: Loudness is closely related to the $L_p$ norm used in existing adversarial attack formulations \eqref{Eq:TAML}. The loudness for each frame is usually calculated as the short-term log-energy \cite{tsai2011automatic}, which is the logarithm of the total energy of the frame. After two short-term log-energy vectors for $s(t)$ and $\hat s(t)$ are obtained, the DTW between them generates the vector of frame-wise deviation values for loudness.  

\end{itemize}

The last step for each feature is to aggregate the computed vector of frame-wise deviation values into a single value to represent the overall feature deviation. According to existing studies \cite{rix2001perceptual,gupta2017perceptual}, the non-linear average calculation is commonly adopted for pitch and rhythm aggregations, and linear averaging is used for timbre and loudness. After the aggregations, the resultant four feature deviation values form a final feature deviation vector to describe the audio characteristic deviation from $s(t)$ to $\hat s(t)$.

\subsection{Impacts of Audio Feature Deviations}
To have a good sense of how pitch, rhythm, timbre, and loudness change in a perturbed music signal, we show the feature deviations caused by an adversarial example in \cite{saadatpanah2020adversarial} in Fig.~\ref{fig:key_insights}.

As \cite{saadatpanah2020adversarial} adopted an $L_p$ norm based formulation to create adversarial audio and limited the $L_p$ norm of the perturbation, Fig.~\ref{fig:waveform} shows that there is a minor waveform change in the time-domain between the original and perturbed music signal. This indicates that the perturbation only incurs a small energy or loudness change to the original signal.

Next, we look at the waveform change in the frequency-domain and compare the power spectrum in Fig.~\ref{fig:Original_music_spectrum}. The observed change is more evident than the time domain in Fig.~\ref{fig:Original_music_spectrum}: the third harmonic in the original harmonics is suppressed, which leads to inharmonicity in the signal and can negatively impact the timbre feature and accordingly the audio quality.

If we look at the pitch contours (i.e., the curves drawn by connecting all pitch values over time) for the original and perturbed signals in Fig.~\ref{fig:Pitch_deviation}, we observe the evident difference of the pitch features between the two signals. Similarly, Fig.~\ref{fig:Rhythm_deviation} shows the optimal DTW path of the perturbed signal to the original one. Intuitively, a music signal with the minimal rhythm deviation should have a nearly straight line DTW path. Fig.~\ref{fig:Rhythm_deviation} shows that the DTW path of the perturbed signal is tortuous compared with the original one.

Note that creating adversarial music inevitably causes some distortions of the original signal. Fig.~\ref{fig:key_insights} demonstrates that there may exist some way to better coordinate such distortions among all audio features to mimic the original signal's quality as much as possible since they are eventually perceived by humans. If we look at the basic adversarial audio attack formulation used in recent research \cite{chen2019real,li2020advpulse,saadatpanah2020adversarial}, the $L_p$ norm of the additive noise is only relevant to the loudness feature without a clear relation to the other three features. It is evident that $L_p$ norm is much easier to compute than pitch, rhythm, and timbre via gradient descend. At the current stage, we do not focus on the computational aspect but on the human perception aspect and continue to understand how these features affect human perception.

\subsection{Human Study Procedures and Setups}\label{SubSec:Human Study Procedures}
\begin{figure}[th]
    \centering
    \includegraphics[width=0.48\textwidth]{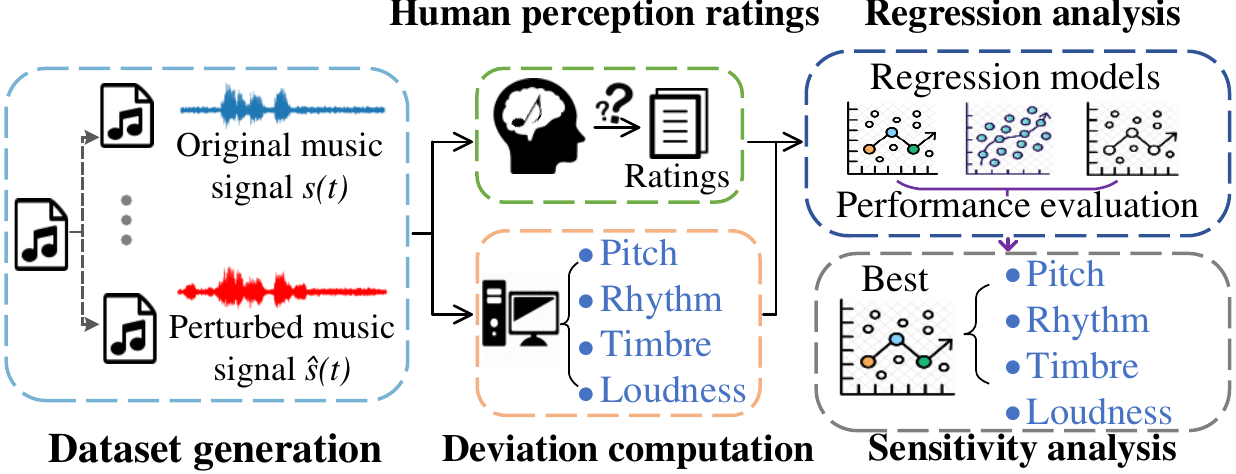}
    \caption{The human study procedure and steps.}
    \label{Fig:Human_study}
\end{figure}

To understand how different features affect human perception. We conduct a human study with the procedure shown in Fig.~\ref{Fig:Human_study}: we first generate a dataset that consists of pairs of original and perturbed music signals. For each pair, we can compute (according to the procedure in Section~\ref{SubSec:Features}) the deviation values for the four features, which form a feature deviation vector. Then, we invite every human participant to assign a deviation rating to each pair based on his/her perceived difference. Next, considering the feature deviation vectors as the inputs and the human ratings as the outputs, we use regression analysis to find the best model to describe the relation between the vectors and the ratings. In this way, we can reverse-engineer the human perception process to build an approximation model to quantitatively predict how much a perturbed signal is perceived by a human.

\begin{figure*}[!t]
    \centering
    \includegraphics[width=0.97\textwidth, trim=4cm 0 4cm 0]{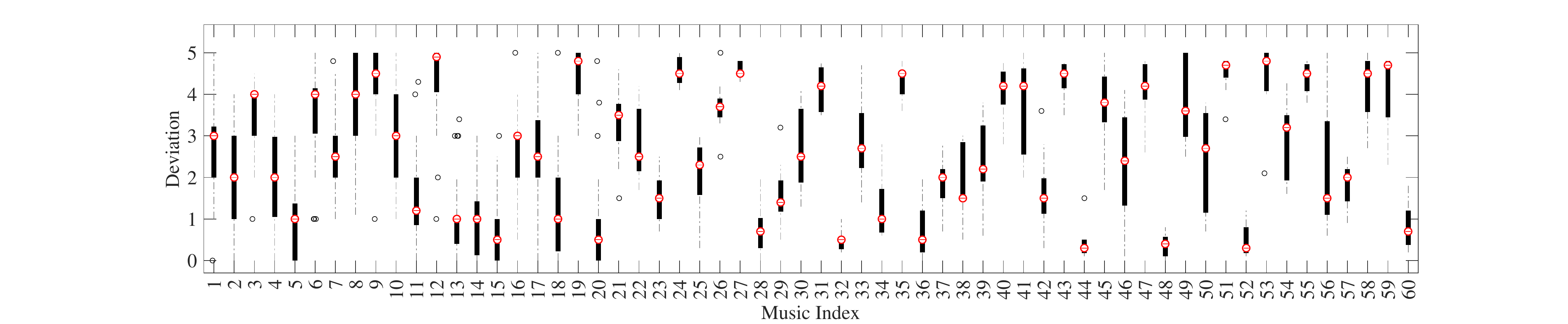}
    \vspace{-0.2cm}
    \caption{Distributions of human ratings of perceived deviation for all pairs of music clips.}
    \label{Fig:boxplot}
\end{figure*}

\textbf{Dataset Generations.}\label{Subsec:dataset}
Since there is no publicly available dataset that provides various versions of perturbed music signals, we propose to generate our own dataset with the following requirements: (i) sufficient diversity of music genres, (ii) sufficient perturbations from the pitch, rhythm, timbre, and loudness perspective, and (iii) slight or moderate perturbation to avoid making participants feel overly noisy.

We build a dataset of 60 pairs of original and perturbed music clips from the genres of Pop, Hip-hop, Rock, Jazz, Classical, R\&B, Country, and Disco. To make participants concentrate on each small perturbation, we crop each music clip to a 5-second WAV format (16kHz, 16-bit PCM, Mono) to avoid audio compression. As there is no guideline or reference to standardize the dataset generation for our study, we aim to create perturbed signals with different feature deviations and varying intensities for human participants such that the data is diverse for regression analysis. Specifically, we use two main mechanisms to create perturbed music clips.


\begin{itemize}
\item Additive noise: an intuitive method is to inject additive noise into the original music. The noise will affect all four features at the same time. To broadly affect the original music, we consider injecting the noise from three aspects: amplitude, frequency and time. To control the amplitude of the noise, we can choose the signal-to-noise (SNR) level from 0dB, 5dB, 10dB, and 15dB \cite{valentini2016investigating}. To inject frequency-sensitive noise, we use both white noise \cite{vaseghi2008advanced} (covering all frequencies with equal intensity) and colored noise (with the power concentrated at certain frequencies). To make noise time-varying, we set random duration and interval of the noise, but the total injection duration is less than the half of the original music length. {\hlb In addition, since existing audio perturbations (e.g., in \cite{saadatpanah2020adversarial}) cause noise-like sounds, the additive noise data rated by human participants should help build a model to properly predict the deviations of noise-like perturbations.}

\item Additive notes: To ensure distinctive deviations among all music features, we also inject additive notes to the original music. To inject notes with the pitch manipulation, we randomly choose notes with the pitch value from 27.5Hz to 4186Hz \cite{lireal} (88 notes space). For rhythm manipulation, we randomly select the additive notes with different lengths and ensure the intervals between adjacent notes are less than 50\% of the original signal's length. To create timbre deviation, we select different instruments to play the additive notes as long as the notes are within the valid pitch ranges of those instruments.
\end{itemize}


\textbf{Human Participation.} We recruited 35 participants \textcolor{black}{who are college students with ages falling between 20 and 35}. All the participants are volunteers without any compensation. Each participant was asked to listen to each pair of the original and perturbed music clips, and then assign a deviation rating on a Likert scale \cite{thiede2000peaq} according to his/her overall music perception: 0$-$1 perfect perceptual quality with imperceptible noise, 1$-$2 good perceptual quality with quiet noise,  2$-$3 noticeable with slight noise,  3$-$4 noticeable and noisy, and 4$-$5 very noisy. More specifically, 1$-$2 means volunteers can only notice some small perturbation after listening to a part of music clips many times, and 2$-$3 indicates the deviation can be noticed by listeners but not noisy. During the experiments, all the volunteers were given the same earphone with the same initial volume setting. They can listen to a music clip as many times as they want.

{\noindent\it Ethical Considerations:} Our study involved human participants that assigned ratings by listening to music. The full protocol was reviewed and exempted by our Institutional Review Board (IRB), which has determined that the study involves the minimal risk for human participants (i.e., the risk is no more than the one that they face during their daily lives). We follow the approved protocol to inform them of the full study procedure and protect their identities without publishing any personally identifiable information.


\textbf{Reverse-Engineering via Regression Analysis.}
Given the computed feature deviations from the original and perturbed music clips as well as the human participant ratings of their perceived deviation, we aim to find the best regression model $M^* \in \mathcal M$ in the model set $\mathcal M$ to minimize the mean squared error (MSE) of regressed prediction, i.e.,
\begin{equation}
     M^* = \underset{M\in \mathcal M}{\arg\min} ~ \mathbb E \| r - M(d_p, d_r, d_t, d_l) \|_2^2,\label{Eq:deviation_weight}
\end{equation}
where $r$ is the human participant rating, $d_p$, $d_r$, $d_t$, and $d_l$ are the deviation values (computed according to the procedure in Section~\ref{SubSec:Features}) for pitch, rhythm, timbre, and loudness, respectively. In our study, we choose Linear Regression \cite{tsai2011automatic,gupta2018technical}, Support Vector Regression, Random Forest, Logistic Regression, and Bayesian Ridge to form the model set $\mathcal M$. With $M^*$ found in \eqref{Eq:deviation_weight}, we use it to quantitatively predict any human-perceived deviation given a pair of original and perturbed music signals.

\subsection{Result Analysis and Discussion}\label{SubSec:Regression Result}
Fig.~\ref{Fig:boxplot} box-plots all the human ratings (ranging from 0 to 5) for individual pairs of music clips from our human study. We can find in Fig.~\ref{Fig:boxplot} that human perception is indeed subjective: each pair of music clips has a range of deviation ratings by different participants; there are always rating outliers for a pair of music clips. Fig.~\ref{Fig:boxplot} also shows that overall, the ratings and the 25\%-75\% boxes are roughly evenly distributed from 0 to 5, which offers sufficient data diversity for regression analysis.

\textbf{Regression Analysis.}
We first use each of Linear Regression, Support Vector Regression (SVR), Random Forest, Logistic Regression, and Bayesian Ridge to model the relationship between feature deviation values and the average human rating, and find the best model with the minimum MSE. We show the MSEs of different regression models during testing in Table~\ref{tab:mse_evaluation}.

\begin{table}[h]
  \caption{MSEs of different regression models.}
  \label{tab:mse_evaluation}
  \vspace{\TableCaptionSpacing}
  \begin{adjustbox}{center,max width=0.95\linewidth}
    \begin{tabular}{cccccc}
      \toprule
      \bf Model: & Linear & SVR      & Random Forest   & Logistic & Bayesian \\
      \midrule
      \bf MSE:   & 1.2351 & 0.8558   & \textbf{0.1541} & 1.6572      & 1.2628    \\
      \bottomrule
    \end{tabular}
  \end{adjustbox}
\end{table}

Through regression analysis, we find that Random Forest performs the best among all the five regression models. As Table~\ref{tab:mse_evaluation} shows, Random Forest leads to an MSE of 0.1541, which is substantially better than Support Vector Regression that achieves the second with an MSE of 0.8558, but an over 5 times increase from Random Forest. The other models result in even worse MSEs. As a result, we choose Random Forest as our regression model to predict the human-perceived deviation. Specifically, given a pair of original and perturbed signals, we name the prediction output of Random Forest as quantified deviation (qDev).



\textbf{Correlation Analysis.}
Then, we analyze to what extent qDev values and realistic human ratings move in tandem; that is, an increase or decrease of value for one will lead to the same for the other. This is important because when creating an adversarial attack against a classifier, we aim to reduce the qDev value of a perturbed signal (so its deviation rating by a human should also decrease) such that the perturbation is hardly noticed by a listener. We use Spearman's rank correlation coefficient \cite{daniel1987spearman,sedgwick2014spearman} to model the correlation in our study. Spearman's coefficient is a commonly used statistic measure to evaluate the relationship between two variables using a monotonic function, where value 1 or -1 indicates that the two always move in the same or opposite direction; value 0 means no correlation.

\begin{table}[h]
  \caption{Spearman's coefficient between the human rating and a deviation measure.}
  \label{tab:Spearman}
  \vspace{\TableCaptionSpacing}
  \begin{adjustbox}{center,max width=\linewidth}
    \begin{tabular}{cccccc}
      \toprule
      \bf Deviation Measure: & {$L_2$} & {$L_{\infty}$} & {SNR} & {qDev} \\
      \midrule
      \bf Spearman's Coefficient:    & 0.3909   & 0.0893   & 0.0134   & \textbf{0.9608}     \\
      \bottomrule
    \end{tabular}
  \end{adjustbox}
\end{table}

Table~\ref{tab:Spearman} lists the Spearman's coefficients between the human rating and each of the following deviation measures: $L_2$ norm \cite{saadatpanah2020adversarial}, $L_{\infty}$ norm \cite{saadatpanah2020adversarial, chen2019real}, SNR \cite{yuan2018commandersong,chen2020devil}, and qDev from Random Forest. It is seen from Table~\ref{tab:Spearman} that qDev has a very high correlation with the realistic human rating, indicating it can be quite useful for predicting a human-perceived deviation of a signal. In other words, minimizing qDev in a mathematical formulation to form an audio signal perturbation would be most likely suppress a human's attention to the signal deviation caused by the perturbation. Interestingly, we also observe that the commonly used $L_p$ norms and SNR are in fact not well related to human perception (e.g. $L_2$ norm has the best correlation of 0.3909). Table~\ref{tab:Spearman} offers quantitative evidence to echo the concern raised in related studies \cite{carlini2017towards, qin2019imperceptible} that suggests new ways to measure the human perceptual similarity may be needed.

\textbf{Sensitivity Analysis.}\label{SubSec:Sensitivity}
To explore which feature is potentially more important than others in human perception, we conduct sensitivity analysis via the One-at-a-time (OAT) strategy \cite{campbell2008photosynthetic, bailis2005mortality, murphy2004quantification}: we remove in turn pitch, rhythm, timbre, and loudness to form three-feature inputs for regression, and measure the MSE of the resultant regression. We find Random Forest is always the best in our OAT analysis to minimize the MSE with only three features reaming as the inputs.

\begin{table}[h]
\centering
\caption{Sensitivity analysis for each feature.}
\label{tab:sensitivity_analysis}
\vspace{\TableCaptionSpacing}
  \begin{adjustbox}{center,max width=\linewidth}
    \begin{tabular}{cccccc}
    \toprule
    Excluding:               & Pitch   & Rhythm  & Timbre & Loudness   & None\\
    \midrule
    MSE:                   & 0.1891 & 0.1581 &0.1889 & 0.3539   & 0.1541   \\
    \bottomrule
    \end{tabular}
  \end{adjustbox}
\end{table}


Table \ref{tab:sensitivity_analysis} shows the MSE of Random Forest for each regression of excluding pitch, rhythm, timbre, and loudness in turn. From Table \ref{tab:sensitivity_analysis}, loudness that represents the energy of the perturbation appears to be the most sensitive feature to human-perceived deviation. For example, removing loudness leads to a 129\% MSE increase from 0.1541 to 0.3539. But it is clear that the other features individually contribute to the overall human perception, and removing one of them causes more MSE in the regression.

Overall, we find in the human study that Random Forest is the best regression model to yield the minimum MSE to predict the human rating as qDev. Simpler regression models, such as Linear Regression or SVR, do not perform as well as Random Forest. This may also confirm that human perception is indeed a complicated process. In addition, qDev is a much more appropriate metric than the conventional $L_p$ norm or SNR in terms of both MSE and Spearman's correlation with the human rating, and the features of pitch, rhythm, timbre, loudness all contribute to the overall perception.

\section{Perception-Aware Attack Strategies}\label{Sec:Attack}
With the metric of qDev regressed via Random Forest from audio features, we reformulate the problem of creating adversarial music signals into a perception-aware attack framework. We then analyze how to narrow down the search space in the reformulation, and eventually find an efficient solution via dynamic clipping.

\subsection{Problem Reformulation}
Existing studies \cite{carlini2018audio,yakura2018robust,li2020advpulse,zheng2021black} solve the original optimization problem in \eqref{Eq:TAML} via finding a sub-optimal yet efficient alternative solution. For our perception-aware reformulation, it is natural to think about reformulating existing alternative solutions by directly replacing its $L_p$ norm with the new metric of qDev. However, such a reformulation no longer offers the advantage of computational efficiency because the process of computing audio features in qDev is unfortunately non-linear, non-convex, and non-differentiable \cite{duan2010multiple}. Accordingly, we formulate the perception-aware attack by replacing $L_p$ norm with qDev in the original form \eqref{Eq:TAML} as
\begin{eqnarray}
 \text{minimize}  &&  \text{qDev}(s(t),\hat{s}(t)),  \label{Eq:deviation}\\
 \text{subject to}  && f(\hat{s}(t)) \neq y, \notag 
\end{eqnarray}
where $\text{qDev}(s(t),\hat{s}(t))$ denotes the qDev between the perturbed signal $\hat s(t) = s(t)+\delta(t)$ and the original one $s(t)$. To ensure $\hat{s}(t)$ to be a valid waveform, we always constrain the normalized amplitude of each of its sample points to be in $[-1,1]$ \cite{chen2019real}.

Finding the optimal solution to \eqref{Eq:deviation} becomes even more difficult than the original one in \eqref{Eq:TAML} because computing qDev involves a much more complicated process than the $L_p$ norm. Our strategy is to analyze what properties the perturbation signal $\delta(t)$ should have towards finding a solution to \eqref{Eq:deviation}.

\subsection{Perturbation Signal Property Analysis}\label{SubSec:Property Analysis}
Since the solution to \eqref{Eq:deviation} is computationally intractable, we have to narrow down the search space for the perturbation signal $\delta(t)$ by analyzing what properties it should have.

The reformulation \eqref{Eq:deviation} means two obvious goals that the perturbation signal $\delta(t)$ should achieve: i) misclassification (i.e., the attack should fool the classifier) and ii) minimized qDev (i.e., it also produces good perceptual quality a human can perceive). At first glance, the two goals seem to contradict with each other (as the best perceptual quality of music indicates no change of its signal and thus no attack success). We need to explore one step further to understand what audio features $\delta(t)$ needs as a result of each of the two goals, then consider all needed features jointly to reconcile any conflict to construct a search space of $\delta(t)$ that is sufficiently narrowed down towards a feasible solution.

\textbf{Properties for Attacking Audio Fingerprinting.}
First, we consider what feature properties $\delta(t)$ should have towards launching a successful attack. A key technique for audio signal classification is audio fingerprinting \cite{cano2005review}. The technique and its variants have been widely adopted in audio signal watermarking \cite{boney1996digital,cox2002digital}, integrity verification \cite{gomez2002mixed}, music information retrieval \cite{wang2003industrial,casey2008content, pardo2006finding}, broadcast monitoring \cite{allamanche2001audioid,neuschmied2001content,haitsma2001robust} and copyright detection \cite{saadatpanah2020adversarial}.

\begin{figure}[!t]
\centering
\begin{minipage}{0.23\textwidth}
  \centering
\includegraphics[width=0.90\textwidth, height=3.28cm]{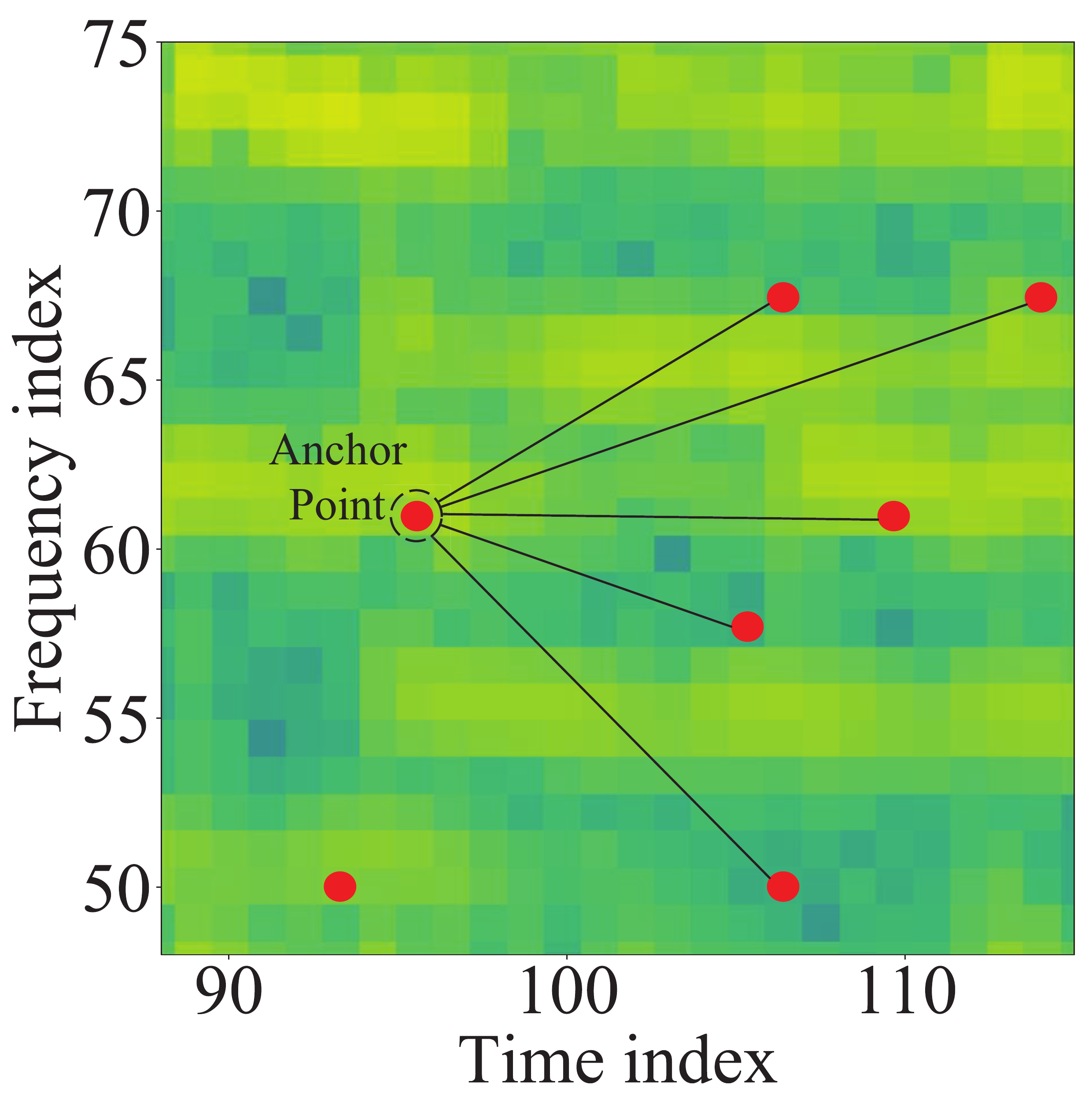}
\subcaption[first caption.]{Fingerprinting generation.}
\end{minipage}%
\hfill
\centering
\begin{minipage}{0.23\textwidth}
  \centering
\includegraphics[width=\textwidth, height=3.3cm]{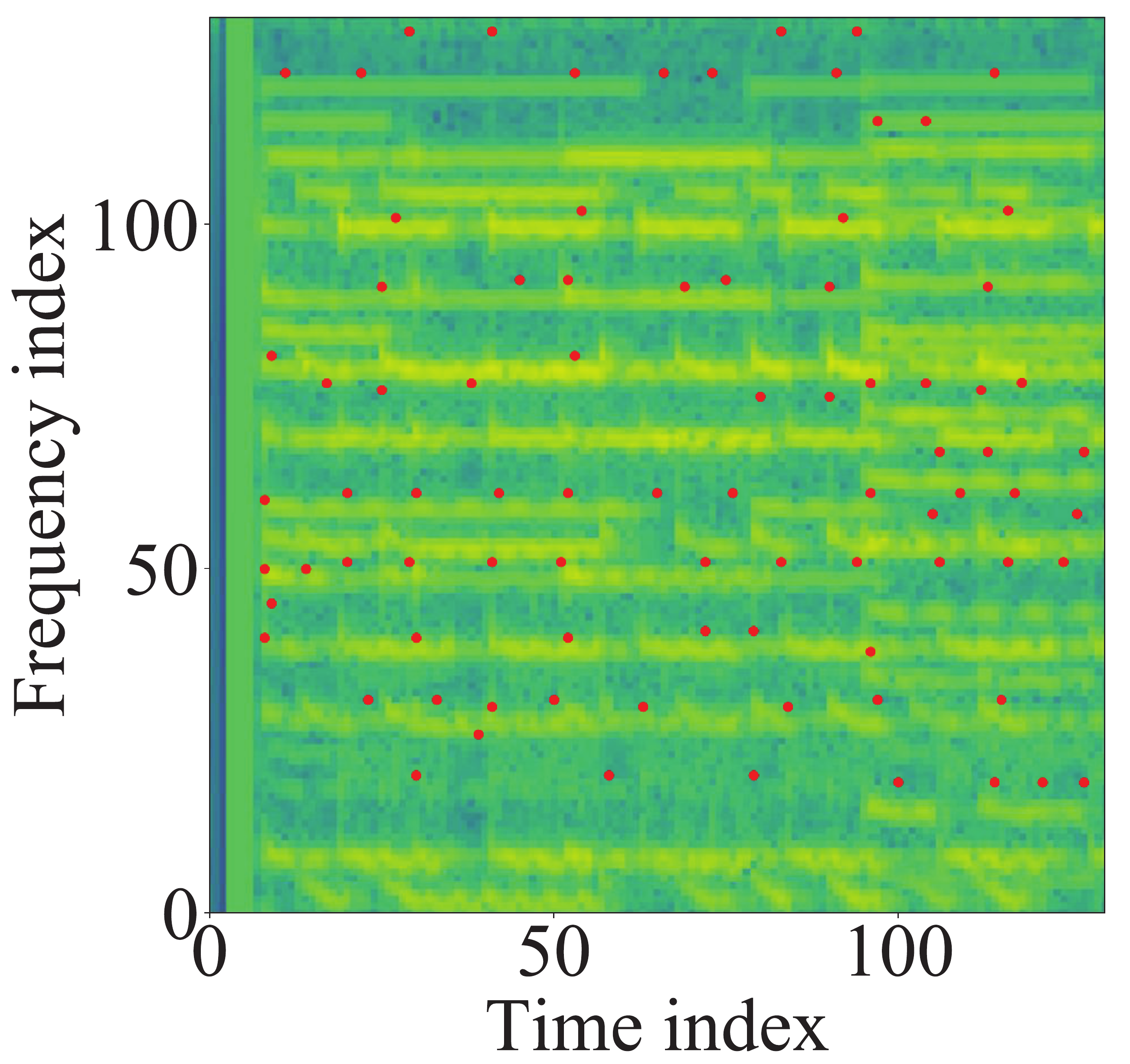}
\subcaption[third caption.]{Distribution of peaks.}
\end{minipage}%
\hfill
\vspace{-0.2cm}
\caption{Fingerprinting generation via finding all peaks in a signal's spectrogram.}
\label{fig:spectrogram}
\end{figure}

The essential idea in audio fingerprinting is to consider certain high-energy areas of an audio signal in the spectrogram as its fingerprints. As an example shown in Fig.~\ref{fig:spectrogram}(a) \cite{wang2003industrial}: an energy peak (anchor point) is paired with other peaks within a certain target area in a signal's spectrogram, then the fingerprints are computed based on the frequency information of the peaks and the time intervals between them. Fig.~\ref{fig:spectrogram}(b) shows there are many peaks in a signal's spectrogram that lead to a large number of fingerprints for audio signal classification and identification.

As we can observe from Fig.~\ref{fig:spectrogram}, peaks in the spectrogram are a key feature for audio signal classification. These peaks are usually the results of a mixture of high-energy points of audio signal harmonics \cite{haitsma2002highly,gomez2002mixed,wang2003industrial}. From the attacker's perspective, creating new positions of harmonics in the spectrogram should be a direct way to manipulate the fingerprints, which can lead to the misclassification of the signal. In the audio features, timbre is the most relevant to the harmonics of the signal \cite{loeffler2006instrument,risset1999exploration}. Given an energy threshold (that represents the loudness) for perturbation $\delta(t)$, a good way to create the attack is to affect the feature of timbre for the signal.

\removecontent{
\begin{table}[h]
\centering
\caption{\hlb{Sensitivity analysis for each two features.}}
\label{tab:sensitivity_analysis_new}
\vspace{\TableCaptionSpacing}
  \begin{adjustbox}{center,max width=\linewidth}
    \begin{tabular}{ccccccc}
    \toprule
    Excluding:         & R\&T   & T\&L  & R\&L & P\&T & P\&L   &P\&R \cr
    \midrule
    MSE:               & 0.3033 & 0.3855 & 0.2071 & 0.3482   & 0.2337 &0.1748   \cr
     \midrule
    Correlation:            & 0.9306 & 0.9189 & 0.9437 & 0.9210   & 0.9363 &0.9575  \\
    \bottomrule
    \end{tabular}
  \end{adjustbox}
\end{table}
}

\textbf{Properties for Good Perceptual Quality.}
Next, we consider what feature properties $\delta(t)$ should have for good music perceptual quality. From the sensitivity analysis in the human study in Section~\ref{SubSec:Sensitivity}, all features, pitch, rhythm, timbre, affect the human perception of signal deviation or the metric of qDev. The change of any of them may result in an increase of qDev and accordingly a noticeable change by human perception. To further explore the relationship between the musical features and human perceived deviations, we remove two features at a time to measure the MSE of the resultant regression. \removecontent{Table~\ref{tab:sensitivity_analysis_new} shows that the MSE increases when we exclude two features at a time. We can see that there is no important features dominating the human-perceived deviation. Hence, considering solution complexity, a reasonable strategy for creating $\delta(t)$ given an energy threshold is to incur the change of only one feature while making the other feature deviations small such that the qDev aggregated over all features remains small. Theoretically, timbre is the only feasible feature that can revise the audio fingerprints without brings deviation to other features \cite{gupta2017perceptual,sethares2005tuning} (i.e., pitch, rhythm, and loudness). On the other hand, changing either pitch or rhythm will revise other feature (i.e., timbre), which is not feasible to preserve the perceptual quality, and it also brings burden to reducing the searching space.}

\textbf{Finding Feasible Search Space.}
To summarize, it would be good to 1) change the feature of timbre for a potentially successful attack, and 2) manipulate only one feature while keeping the others unchanged as much as possible to maintain the perceptual quality. \removecontent{Changing pitch or rhythm can also revise the fingerprints, but it is not feasible to solve the perception problem as we discussed above. Hence,} to reconcile the two requirements: we propose to change timbre much more than the other features.

Now the question becomes how to create $\delta(t)$ with a quite different timbre feature while maintaining almost the same pitch and rhythm features. The traditional perturbation design in \eqref{Eq:TAML} usually generates a noise-like perturbation and is not able to create this required signal because it causes all distortions of pitch, rhythms, and timbre (as shown in Fig.~\ref{fig:key_insights}). 
As a music signal consists of well-crafted, human-enjoyable musical notes, we propose to create $\delta(t)$ by reproducing the same music notes via new instruments. The timbre feature is always associated with the harmonics, and we can find these natural harmonics from the instruments. In this way, the timbre of $\delta(t)$ can be changed substantially due to different harmonic characteristics of new instruments; but pitch and rhythm may deviate less if we find appropriate instruments to play the same notes. To demonstrate the feasibility of our design, we compare the feature deviations of a perturbed music signal mixed by randomly-generated noise and instrument-generated music notes.


\begin{table}[h]
  \caption{Noise vs notes played by a different instrument.}
  \label{tab:ins_vs_noise}
  \vspace{\TableCaptionSpacing}
  \begin{adjustbox}{center,max width=\linewidth}
    \begin{tabular}{lcccccccc}
      \toprule
      \bf ~
      & Pitch
      & Rhythm
      & Timbre
      & Loudness
      & qDev\\
      \midrule
      \bf Instrument:     & 0    & 0.85    & 25320   & 2873  & 2.23\\
      \midrule
      \bf Noise:          &0.9049 &7.239   &19521    &1988   &3.86 \\
      \bottomrule
    \end{tabular}
  \end{adjustbox}
\end{table}


As shown in Table~\ref{tab:ins_vs_noise}, the additive instrument produces a higher loudness value than noise (indicating a more energy level); at the same time, it generates more timbre deviations (25320 vs 19521) but less pitch and rhythm deviations than the noise. Depending on the difference between $s(t)$ to $\hat s(t)$, the non-linearly aggregated pitch and rhythm deviations have values commonly in the range from 0 to 50, and the linearly aggregated timbre and loudness deviations usually range from zero to tens of thousands. There exists an obvious deviation gap between instrument-generated notes and randomly-generated noise of different features. We also use qDev to quantify the deviations, and the instrument-generated notes have a clearly lower qDev value than random noise (2.23 vs 3.86). This makes it a much more desirable signal component for $\delta(t)$ in terms of both human-perceived quality (low qDev) and attack effectiveness (more timbre variation).

Consequently, we can effectively narrow down the search space by considering $\delta(t)$ as a linear combination of signals consisting of the same music notes played by different instruments for the original music signal. Then, generating the perturbed signal $\hat s(t) \!=\! s(t) \!+\! \delta(t)$ is like finding ``subtle'' instrumental track signals then optimally remixing them (based on qDev) into the original music.

It is worth mentioning that a music signal can consist of both instrumental and vocal tracks. It is possible to add a new vocal track (i.e., the same vocal notes sung by a different voice to change the feature of timbre) into the perturbation $\delta(t)$. As it is easier to generate instrumental signals by computer music synthesis, we only use instrumental tracks to form $\delta(t)$ in this paper. \removecontent{However, we tested the additive notes based music clips in Section~\ref{Subsec:dataset} were not effective to spoof the copyright detector, because the purpose of generating music clips is to better understand human perception to different perturbation level rather than considering bypassing detector. Therefore, we formulate a specific attack with considering both attack effectiveness and perceptual quality in the following section.}


\subsection{Perception-Aware Attack Formulation}
With the shrunk search space, we write $\delta(t)=\sum_{k=1}^K \theta_k \delta_k(t)$, where $K$ denotes the number of different instrumental tracks, $\delta_k(t)$ is the $k$-th instrumental track signal, and $\theta_k$ is the non-negative weight for $\delta_k(t)$. Next, we reformulate \eqref{Eq:deviation} into a perception-aware attack of finding the best linear weights $\theta_k$ in $\delta(t)$ to minimize the qDev:
\begin{eqnarray}
 \underset{\{\theta_k\}_{k\in[1,K]}}{\text{minimize}}  &&  \text{qDev}\! \left(s(t),s(t)+ \sum\nolimits_{k=1}^K \!\theta_k \delta_k(t)\right)  \label{Eq:deviation2}\\
 \text{subject to}
 && f\! \left(s(t)+\sum\nolimits_{k=1}^K \!\theta_k \delta_k(t)\right) \neq y,\label{Eq:flag2} \notag \\
 && \sum\nolimits_{k=1}^{K} \!\theta_k = \epsilon, \label{Eq:loudness_threshold2}\\
 && \mathcal{P}_{s(t)} \!\subseteq\!  \mathcal{P}_{\delta_k(t)} \, \forall \, k\!\in\! \{k|k \!\in\![1,\!K], \theta_k \!\neq\! 0\}, \label{Eq:pitch_range}
\end{eqnarray}
where \eqref{Eq:loudness_threshold2} ensures the energy level of the perturbation signal $\delta(t)$ is less than a threshold $\epsilon$, $\mathcal{P}_{s(t)}$ and $\mathcal{P}_{\delta_k(t)}$ in \eqref{Eq:pitch_range} represent the sets of pitch values in the original signal $s(t)$ and the $k$-th track signal $\delta_k(t)$, respectively; \eqref{Eq:pitch_range} ensures that $\delta_k(t)$ covers the pitch range of $s(t)$ so the pitch feature of $\delta_k(t)$ does not deviate much from $s(t)$.

The optimization \eqref{Eq:deviation2} is a problem of finding the optimal linear weights. Although still non-differentiable, \eqref{Eq:deviation2} opens a door for a grid search based heuristic solution. Specifically, we can let each linear weight $\theta_k$ be a multiple of a small step $\Delta$ (that is a fraction of the threshold $\epsilon$ in \eqref{Eq:loudness_threshold2}), then enumerate all combinations of possible values for $\{\theta_k\}_{k\in[1,K]}$ to find a solution to \eqref{Eq:deviation2}. For example, setting $\Delta \!=\! 0.1 \epsilon$ and $K \!=\! 10$ produces 92,378 combinations in total. Iterating through them, though not very efficient, is quite feasible for an attacker's computing capability today.

\subsection{Dynamic Clipping}\label{SubSec:DynamicClipping}
\begin{figure}[!t]
    \centering
    \includegraphics[width=0.4\textwidth]{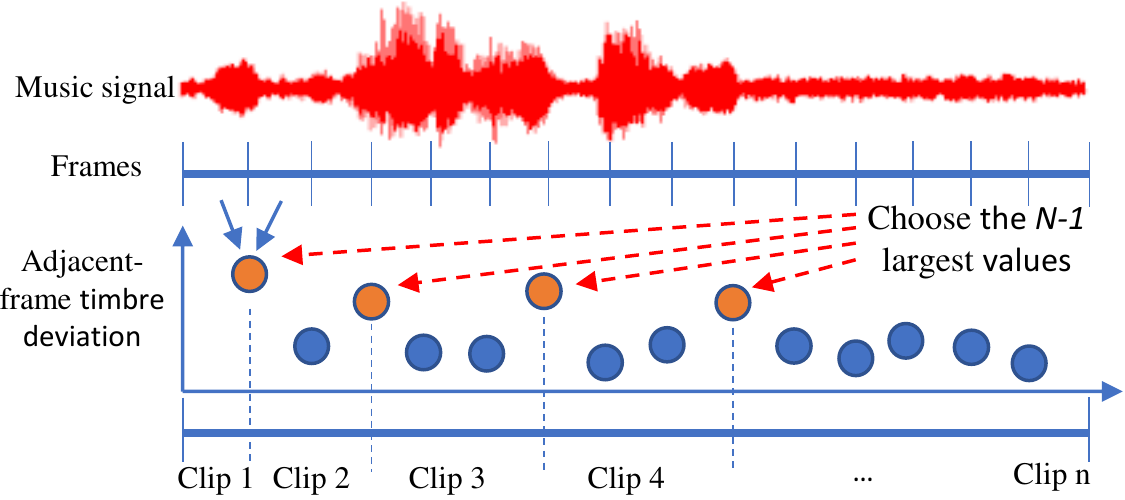}
    \caption{Overview of dynamic clipping.}
    \label{Fig:clipping}
\end{figure}

The optimization in \eqref{Eq:deviation2} finds out a perturbation signal $\delta(t)$ based on the entire duration of the original signal $s(t)$. However, a piece of music can consist of multiple segments with audio characteristics varying within a wide range of instruments and vocals, creating distinct timbre features. For better perceptual quality and attack effectiveness, it is necessary to segment $s(t)$ into $N$ clips according to evident timbre changes and create the perturbation for each clip using the clip-wise optimization based on \eqref{Eq:deviation2}. We call this procedure dynamic clipping.

Fig.~\ref{Fig:clipping} shows the overall process of dynamic clipping: in order to dynamically segment $s(t)$ into $N$ clips, we first separate $s(t)$ into small frames and compute the timbre deviation between each pair of adjacent frames (using the timbre deviation calculation discussed in Section~\ref{SubSec:Features}). Then, we identify $N\!-\!1$ pairs which have the $N\!-\!1$ largest adjacent-frame deviation values, as they contain the most evident $N\!-\!1$ changes of timbre over the duration of the music. We use the timing boundary between two frames in a pair as a timing position to segment $s(t)$. In this way, $s(t)$ is segmented into $N$ clips, each of which will be used to find a corresponding perturbation based on \eqref{Eq:deviation2}.

\section{Realistic Black-box Attack against Copyright Detector}\label{Sec:RealAttack}
In this section, we create a realistic attack based on the perception-aware attack framework in Section~\ref{Sec:Attack}. We choose the YouTube copyright detector as our target as YouTube has exhibited some robustness against noise and perturbations \cite{saadatpanah2020adversarial}. Because there is no knowledge of YouTube's design, we create our own detector based on open-source information for an adversarial transfer attack. We first present how to generate additional instrumental tracks for the perturbation signal given a music signal, then describe the design of our detector as a surrogate model for YouTube's detector.

\subsection{Perturbation Signal Generation}
Perturbation signals generated by \eqref{Eq:deviation2} require the detailed music notes of the original music. For a popular piece of music, its Musical Instrument Digital Interface (MIDI) file is usually available in online databases (e.g., FreeMidi.org and Nonstop2k\footnote{FreeMidi.org:https://freemidi.org/,  Nonstop2k:https://www.nonstop2k.com/}). The MIDI file contains all instrumental tracks with music notes. We use Music21$\footnote{Music21 is a Python-based toolkit for computer-aided musicology. In this work, we use it to produce different instrumental tracks playing the same musical notes}$ to play a downloaded MIDI file with different instruments to form a perturbation for \eqref{Eq:deviation2}.
To achieve the diversity of the timbre feature for \eqref{Eq:deviation2},  we consider an instrument set of instruments across the four families
    \textit{stringed} (Guitar, Electric Guitar, Violin, Viola, Cello, Bass, Electric Bass),
    \textit{woodwind} (Clarinet, Flute, Saxophone, Oboe, Bassoon),
    \textit{brass} (Trumpet, Baritone, Tuba, Horn, Trombone),
    \textit{keyboard} (Piano, Electric Piano).
We empirically select at most two instruments from each family based on a music genre to reduce the computational complexity and the pitch range requirement for perturbation generation in \eqref{Eq:pitch_range}.

\subsection{Surrogate Detector}
\textbf{Audio Fingerprints.} A copyright detector takes audio fingerprinting features as the input. We select the fingerprints and their extraction method introduced in \cite{wang2003industrial}. We extract fingerprints by considering the time, frequency, and amplitude data of the audio. Specifically, we use Fast Fourier Transform (FFT) to generate a spectrogram of an audio signal and extract the spectral peaks of acoustic harmonics, which are shown invariant and reproducible from signal degradation \cite{cano2002robust} and robust to noise and distortion \cite{wang2003industrial}. We then apply the fast combinatorial hashing method \cite{wang2003industrial} to form these fingerprints to hashes for the similarity comparison later.

\textbf{Detection Design.}\label{SubSec:DetectorDesign}
The detection is built to compute the similarity of the fingerprints of an input signal to the detector's database. If the similarity score is higher than a similarity threshold, the detector will raise an alarm. To ensure our surrogate detector has a degree of transferability to YouTube's detector, we must adopt a threshold that is similar to YouTube's. We note that our objective is not to precisely rebuild YouTube's model, but to choose an appropriate threshold (even in a rough way) such that we can use the surrogate detector to predict the output label during minimizing qDev in \eqref{Eq:deviation2}. Because music consists of diversities of audio features, we choose one threshold for each of 8 music genres: Pop, Hip-hop, Rock, Classical, Jazz, R\&B, Country, and Disco.

\begin{figure}[!t]
    \centering
    \includegraphics[width=0.5\textwidth]{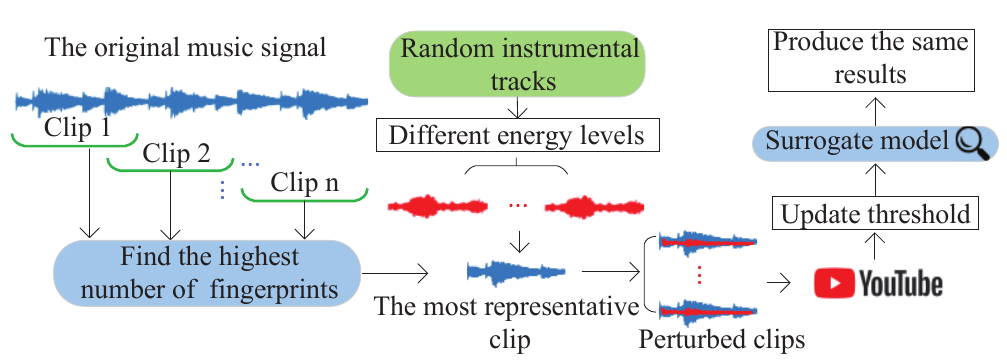}
    \caption{Process of obtaining the threshold from YouTube.}
    \label{Fig:Interaction}
\end{figure}

Fig.~\ref{Fig:Interaction} shows the process we use to approximately calibrate the surrogate detector's threshold towards YouTube's. This process is similar to the one proposed in \cite{chen2019real} that estimates the threshold of a black-box model. In particular, to obtain the threshold for a music genre, we choose a song from the genre, crop it into clips, choose the most representative clip that contains the highest number of fingerprints among all the clips. Then, we randomly add instrumental track signals with different energy levels  to this clip, generating a number of clips with perturbations of varying energy levels. We send these clips to YouTube to see the copyright detection results, and set the detection threshold for the surrogate detector such that it yields the same results as YouTube does.

\begin{figure*}[!t]
  \centering
  \includegraphics[width=0.97\textwidth]{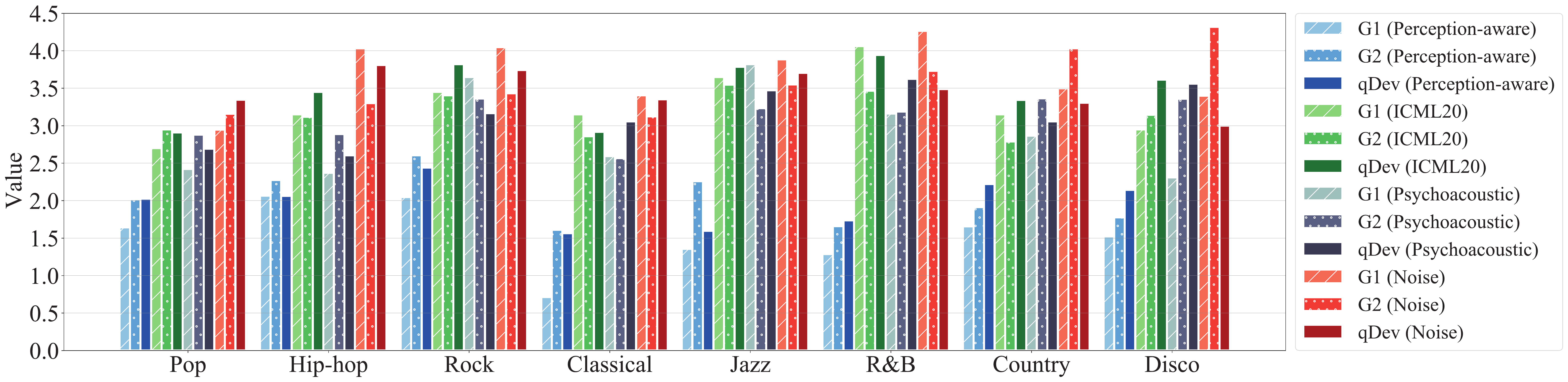}
  \vspace{-0.3cm}
  \caption{\hlb{Human ratings and qDev values: Perception-aware, ICML20, psychoacoustic, and random noise attacks.}}
  \label{Fig:Compare_listening}
\end{figure*}

\section{Experiments and Results}\label{Sec:Experiments}
In this section, we present the experiments and results. We first describe the experimental settings, then discuss the audio perceptual quality and attack effectiveness of generated adversarial music.

\subsection{Experiments Setup}\label{sec:experiment_setup}
\noindent\textbf{Music Dataset:}
To cover a wide range of music data, we selected 32 top hits songs of the last 20 years from 8 genres: Pop, Hip-hop, Rock, Classical, Jazz, R\&B, Country, and Disco. We created \textcolor{black}{56 clips} of 5--10 seconds and 16 clips of 30 seconds clips for human evaluation, and 160 clips of 30 seconds for attack strength evaluation. We have verified that all the clips were copyright-detected by YouTube.


\noindent\textbf{Default Experimental Setups:}
The default settings in \eqref{Eq:deviation2} for the perception-aware attack include the search step $\Delta\!=\!0.1\epsilon$, the number of instruments for perturbation generation $K\!=\!7$, and the number of clips in dynamic clipping $N\!=\!6$.

\noindent\textbf{Attack Method Comparison:}
\textcolor{black}{We compare the perception-aware attack with two recent attack methods: the ICML20 method against YouTube in \cite{saadatpanah2020adversarial} and the psychoacoustic attack framework \cite{schonherr2018adversarial,qin2019imperceptible,lireal}. Specifically, for the ICML20 attack, we directly adopted the source code provided by the authors of \cite{saadatpanah2020adversarial}; for the psychoacoustic attack, we followed the two stage attack introduced in \cite{qin2019imperceptible}: we first used the ICML20 method to generate an adversarial music perturbation then applied the iterative process that involves the masking threshold \cite{qin2019imperceptible} instead of the $L_p$ norm in the loss function to improve the perception.} Finally, We implemented a random noise attack method that adds random noise to music as a baseline case.

\textcolor{black}{Here we provide a YouTube link that demonstrates adversarial clips created by the perception-aware attack in comparison with other attacks: \url{https://www.youtube.com/watch?v=IfBAzmdN5ds}}.


%
%
%

\subsection{Perceptual Quality of Adversarial Music}\label{suc:human_evaluation}
We first evaluate the perceptual quality of adversarial music created by the perception-aware, ICML20, \textcolor{black}{psychoacoustic attack}, and random noise attacks. In the experiments, given original music, we created perturbed music clips of 5--10 seconds under each attack by increasing the energy threshold of the perturbation such that the perturbed clip exactly bypassed YouTube's detector. For each perturbed clip, we used the Random Forest regressed qDev in Section~\ref{SubSec:Regression Result} to predict its deviation from the original clip.

{\hlb
\noindent\textbf{Human Evaluations:}
We involved 14 of 35 human volunteers in the training study in Section~\ref{Sec:Human Perception} to participate the evaluations. They formed Group~1 (G1) in our evaluations. We also recruited additional 15 college student volunteers, referred to as Group~2 (G2), to participate the evaluations. The age ranges of G1 and G2 are 20-34 and 22-33, respectively. The results of G1 can show the test accuracy of the regressed qDev model and the results of G2 can further demonstrate the generalizability of the regressed model (i.e., the training model built from a group of people can be used to predict the rating of another group of new people who are not in the training set). Every music clip in our evaluations was rated by all participants in both G1 and G2.}

\noindent\textbf{Human Rating vs qDev under Different Attacks:}
Fig.~\ref{Fig:Compare_listening} illustrates the average human ratings and qDev values of the perception-aware, ICML20, \textcolor{black}{psychoacoustic}, and random noise attacks for each music genre. It is evident from the figure that the perception-aware attack always achieves much smaller deviation ratings and qDev values than the other \textcolor{black}{three} attacks. For example, for classical music, the perception-aware attack obtains ratings of 0.71 (G1) and \textcolor{black}{1.61 (G2) (indicating perfect and good perceptual quality according to the rating guideline, respectively)} while the ICML20, \textcolor{black}{psychoacoustic, and noise attacks get 3.15 (G1) vs 2.91 (G1), 2.59 (G1) vs 2.56 (G2), and 3.40 (G1) vs 3.12 (G2), respectively (indicating noticeable and noisy).} It is also observed that rock music seems harder to perturb for the perception-aware attack and has ratings \textcolor{black}{2.04 (G1) and 2.60 (G2)} (noticeable with slight noise). 
Overall, Fig.~\ref{Fig:Compare_listening} shows that the perception-aware attack achieves substantially better perceptual quality than the ICML20, \textcolor{black}{psychoacoustic}, and random noise attacks.

{\hlb
\noindent\textbf{Accuracy of qDev-based Prediction:}
By comparing the qDev value with the human rating in every genre in Fig.~\ref{Fig:Compare_listening}, we can see that qDev is a good prediction to the human rating as the qDev does not deviate much from the average human rating for each genre. For example, the Hip-hop music created by the perception-aware attack has the qDev of 2.06 compared with the average human rating of 2.07 (G1) and 2.27 (G2). Table~\ref{tab:MSE_qDev_genres} shows the MSE between the qDev value and the average G1 rating compared with the MSE between the qDev value and the average G2 rating in each genre. The MSEs averaged over all genres are 0.4107 (G1) and 0.5848 (G2), which are both higher than the training MSE of 0.1541 in Table~\ref{tab:mse_evaluation} in Section~\ref{SubSec:Regression Result}. Overall, it is observed that G2 incurs a slightly higher average MSE than G1 in the evaluation as new participants bring new subjective judgements.

\begin{table}[t]
  \caption{\hlb{The MSEs of qDev among genres (G1 vs G2).}}
  \label{tab:MSE_qDev_genres}
  \vspace{\TableCaptionSpacing}
  \begin{adjustbox}{center,max width=\linewidth}
    \begin{tabular}{cccc}
      \toprule
      Pop & Hip-hop & Rock & Classical \\
      \midrule
      0.726 vs 1.027  & 0.161 vs 0.316 & 0.235 vs 0.382  & 0.480 vs 0.230 \\
      \midrule
      Jazz & R\&B & Country & Disco \\
      \midrule
      0.153 vs 0.333 & 0.426 vs 1.277 & 0.444 vs 0.299 & 0.735 vs 1.070 \\
      \bottomrule
    \end{tabular}
  \end{adjustbox}
\end{table}

Table~\ref{tab:mse_evaluation_regression} compares the MSEs for G1 and G2 in different regression models built from the training in Section~\ref{SubSec:Regression Result}. We can see that all the test MSEs increase from the training MSEs in Table~\ref{tab:mse_evaluation}, and Random Forest still achieves the minimum MSE for both G1 and G2. In all regression models, Random Forest also exhibits the minimum MSE increases from the training MSE to the test MSEs of either G1 or G2.
}

\begin{table}[h]
  \caption{{\hlb MSEs of different regression models.}}
  \label{tab:mse_evaluation_regression}
  \vspace{\TableCaptionSpacing}
  \begin{adjustbox}{center,max width=0.95\linewidth}
    \begin{tabular}{rccccc}
      \toprule
      \bf Model: & Linear & SVR      & Random Forest   & Logistic & Bayesian \\
      \midrule
      \bf MSE-G1:   & 1.5826 & 2.2894   & \textbf{0.4107} & 1.9263      & 1.5012    \\
      \midrule
      \bf MSE-G2:   & 1.9568 & 2.6169  & \textbf{0.5848} & 2.2103      & 1.8521    \\
      \bottomrule
    \end{tabular}
  \end{adjustbox}
\end{table}

{\hlb
\noindent\textbf{Role of Additive Noise Data in Training:} As we discuss previously, to evaluate the importance of additive noise data for building an accurate regression model in Section~\ref{Subsec:dataset}, we only use the additive note data to train a new qDev$^*$ metric and compare the prediction of qDev$^*$ with G1 and G2 ratings. Specifically, it is observed that when we replace qDev with qDev$^*$, the MSE increases from 0.4107 to 2.0054 for G1, and from 0.5848 to 2.2944 for G2. As a result, additive noise is essential to build an accurate qDev model.}

\noindent\textbf{Impact of Dynamic Clipping:}
We also evaluate the impact of dynamic clipping in Section~\ref{SubSec:DynamicClipping} on the overall perceptual quality of the perturbed music. We compare its performance with a static clipping design in which a clip is uniformly segmented into 6 smaller clips with equal length for perturbation generation.

Table~\ref{tab:remixing_evaluation} shows the qDev values of the two designs for different music genres. We can observe that dynamic clipping achieves uniformly better perceptual quality in all genres.

\begin{table}[h]
  \caption{qDev values in dynamic vs static clipping.}
  \label{tab:remixing_evaluation}
  \vspace{\TableCaptionSpacing}
  \begin{adjustbox}{center,max width=\linewidth}
    \begin{tabular}{lcccccccc}
      \toprule
      \bf ~
      & Pop
      & Hip-hop
      & Rock
      & Classical\\
      \midrule
      \bf Dynamic:     & 1.8953    & 2.9250    & 2.6051   & 1.4956 \\
      \bf Static: & 2.2522    & 3.1854    & 3.1955   & 1.7558 \\
      \midrule
      \bf ~
      & Jazz
      & R\&B
      & Country
      & Disco\\
      \midrule
      \bf Dynamic:     & 1.8653   & 1.3897  & 1.6925  &2.1933   \\
      \bf Static:      & 2.9192   & 2.0925  & 2.0230  &2.2588   \\
      \bottomrule
    \end{tabular}
  \end{adjustbox}
\end{table}



\noindent\textbf{Perceptual Quality without Reference.} Previous experiments were conducted in a formal lab setting to quantify the perceived deviation via actual human ratings and qDev estimates. When a person listens to music during the daily life, there is no reference for him/her to perceive a deviation. The person may or may not notice an issue if the music is perturbed.

We conducted another experiment to measure how human participants perceive perturbed music without reference. In particular, we selected 16 30-second music clips, and asked two questions to each participant for each clip: (i) If familiar with the music: Assign a deviation rating based on your memory using the same rating guideline. (ii) Otherwise: Do you feel abnormal about the music? Please answer 1) Yes, 2) No, or 3) Not Sure. {\hlb We totally received 140 ratings in (i) and 84 answers in (ii) from G1, and 162 ratings in (i) and 78 answers in (ii) from G2.}



Table~\ref{Tab:Raw_listening_familiar} shows the average human ratings \textcolor{black}{along with the number of ratings received in each genre} without reference and average qDev values for different music genres. We can find that the rating distribution among music genres is quite similar to Fig.~\ref{Fig:Compare_listening}. For example, the Classical music can still achieve nearly perfect perceptual quality of \textcolor{black}{0.86 (G1) and 1.73 (G2)}. Rock and Hip-Hop are the worst genres to perturb and make human participants feel noticeable with slight noise deviations. Interestingly, \textcolor{black}{we find that the human rating of G1 and G2 for R\&B music is 0.5 (nearly perfect) and 1.43 (good quality) without reference, which are both improved from the experiments with reference. The potential reason is that the additive instrumental track signals sound natural and embedded to the original music. It becomes hard for humans to recognize these timbre changes without reference. Overall, the G1 and G2 ratings are very similar in Pop, Hip-hop, Rock, and Jazz; and the ratings averaged over all genres are also close (G1: 1.62 vs G2: 2.12).}


\begin{table}[h]
  \caption{\hlb{Human ratings without reference and qDev.}}
  \label{Tab:Raw_listening_familiar}
  \vspace{\TableCaptionSpacing}
  \begin{adjustbox}{center,max width=0.85\linewidth}
    \begin{tabular}{lcccccccc}
      \toprule
      \bf ~
      & Pop
      & Hip-hop
      & Rock
      & Classical\\
      \midrule
      \bf G1 rating (140):     & 1.4500 (18)    & 2.4428 (14)    & 2.4867 (21)   & 0.8583 (12) \\
      \bf G2 rating (162):     & 1.9993 (24)    & 2.6408 (22)    & 2.7988 (22)   & 1.7367 (14) \\
      \bf qDev:             & 1.7850    & 2.7133    & 2.5653   & 1.6255 \\
      \midrule
      \bf ~
      & Jazz
      & R\&B
      & Country
      & Disco\\
      \midrule
      \bf G1 rating (140):     & 1.7500 (21)   & 0.5000 (20)  & 1.4458 (16)  &1.4821 (18)   \\
      \bf G2 rating (162):     & 1.8909 (22)    & 1.4333 (20)    & 2.6606 (18)   & 2.0053 (20) \\
      \bf qDev:             & 2.5679   & 1.4905  & 1.6925  &2.1178   \\
      \bottomrule
    \end{tabular}
  \end{adjustbox}
\end{table}

\begin{figure}[!t]
\centering
\begin{subfigure}{0.5\columnwidth}
\includegraphics[width=\columnwidth]{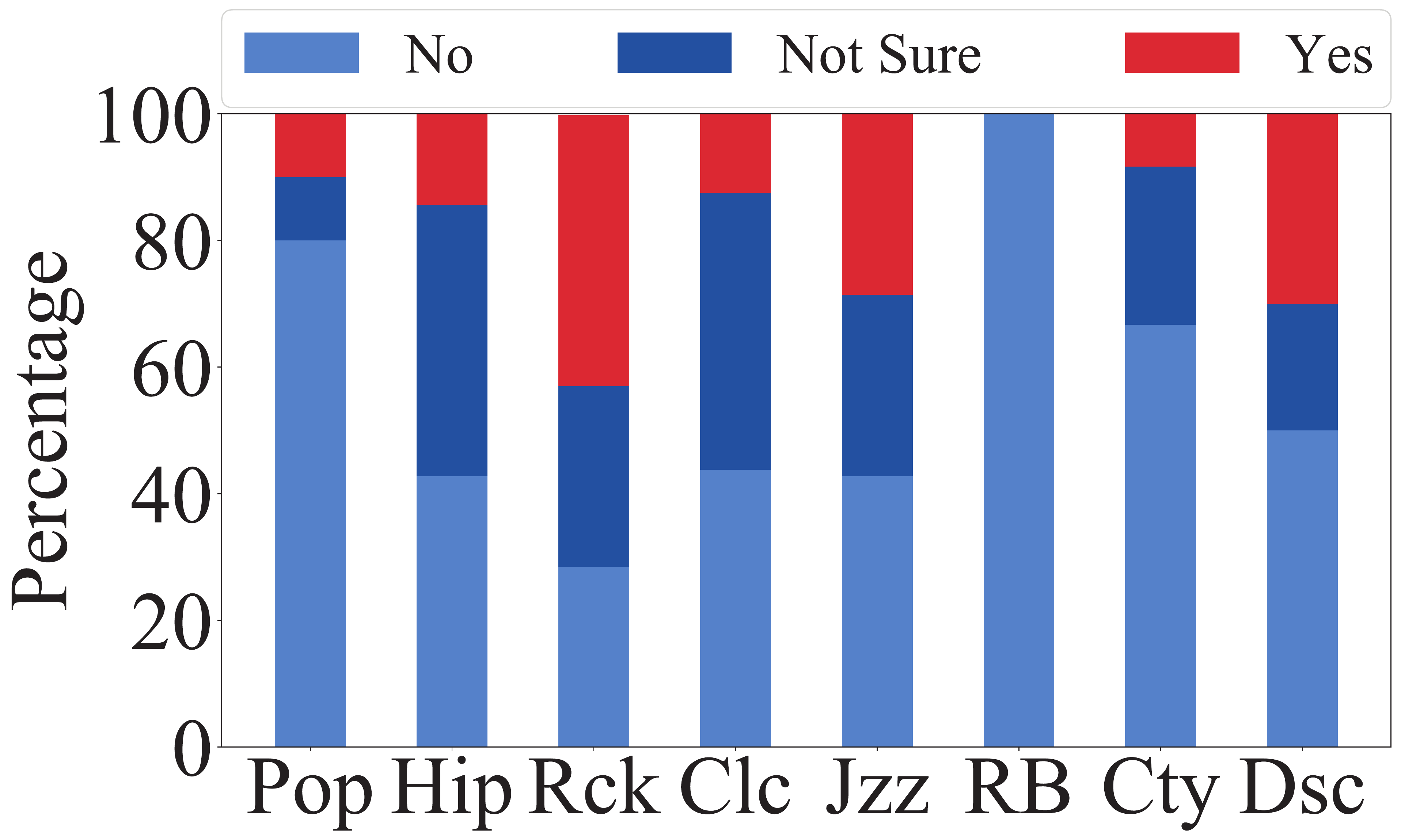}
\caption{Group1}\label{genre_deviation1}
\end{subfigure}%
\hfill
\begin{subfigure}{0.5\columnwidth}
\includegraphics[width=0.90\columnwidth]{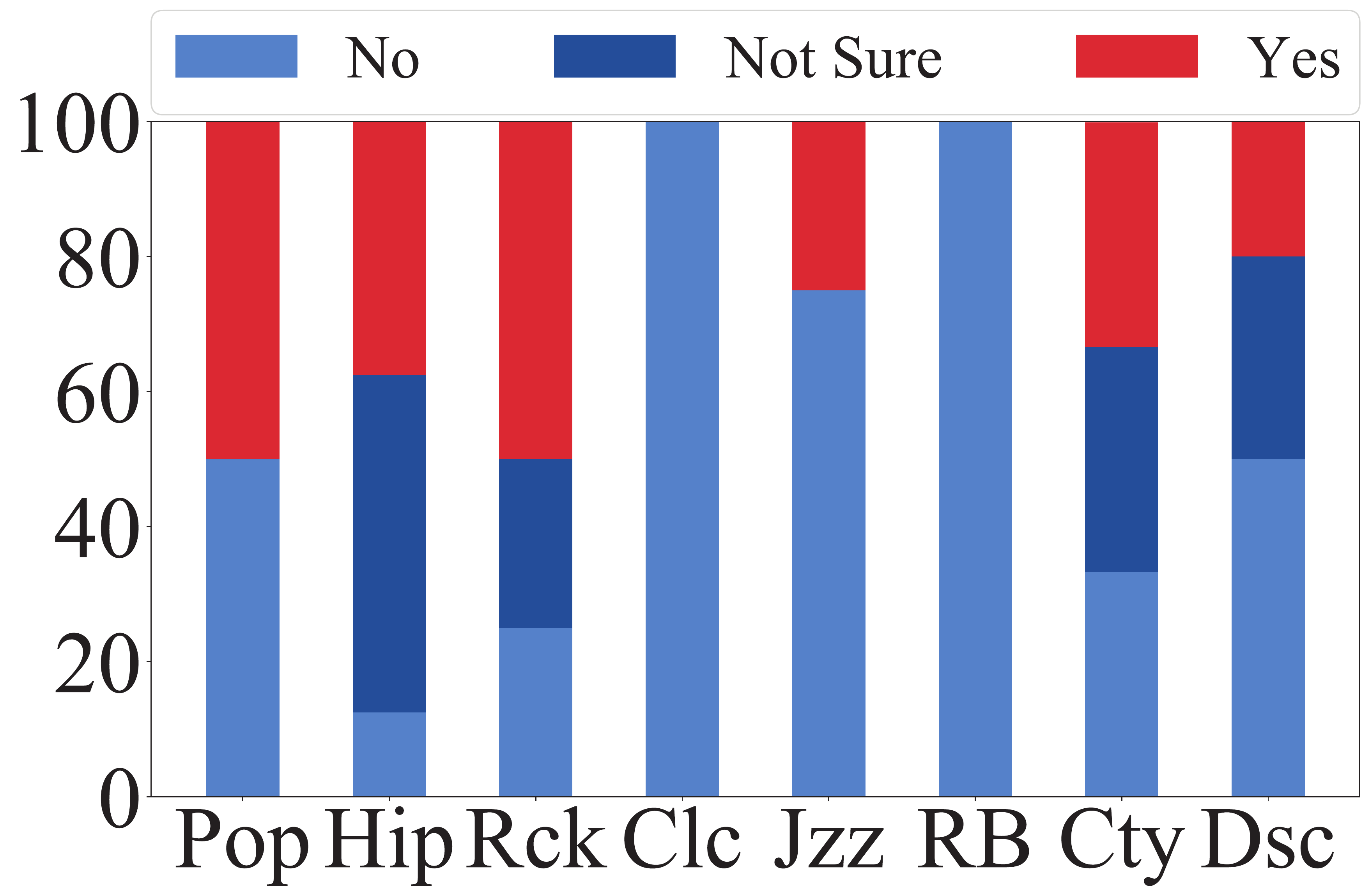}
\caption{Group2}\label{genre_deviation2}
\end{subfigure}%
\hfill
\vspace{-0.2cm}
\caption{\hlb{Percentages of answers by participants of different groups unfamiliar with the given music. The numbers of answers received are 10, 14, 7, 16, 7, 8, 12, 10 in G1 (total: 84), and 6, 8, 8, 16, 8, 10, 12, 10 in G2 (total: 78) for Pop, Hip-hop, Rock, Classical, Jazz, R\&B, Country, and Disco, respectively. }}
\vspace{-0.1cm}
\label{Fig:Raw_listening_unfamiliar}
\end{figure}




Fig.~\ref{Fig:Raw_listening_unfamiliar} depicts a more interesting result of the percentages of different answers by participants unfamiliar with the given music. {\hlb We can see that most participants do not surely notice any abnormality in perturbed soft music (e.g., R\&B and Classical). For example, no audience finds any issue in any R\&B music for both G1 and G2; and 30\% or more answers for Rock music clips are abnormal. Fig.~\ref{Fig:Raw_listening_unfamiliar} shows that the majority of participants (i.e., 81.25\% in G1 and 69.70\% in G2) do not clearly notice the music perturbations generated by the perception-aware attack}. Considering the fact that participants may form a cognitive bias in the study (i.e., they might feel ``obliged'' or ``mentally-focused'' to identify an abnormality), we think that a casual listener without reference might be more unlikely to notice the perturbation of adversarial music created by the perception-aware attack.

\subsection{Attack Effectiveness vs qDev}
Next, we measure the attack success rates of the perception-aware, ICML20, \textcolor{black}{psychoacoustic}, and random noise attacks against YouTube. As discussed in Section~\ref{SubSec:DetectorDesign}, the fingerprinting similarity thresholds in our surrogate detector were set roughly according to YouTube's detection results using a few music samples. But an adversarial music clip bypassing the surrogate detector does not necessarily mean that it will also evade YouTube's detection. In this experiment, we used the perception-aware, ICML20, \textcolor{black}{psychoacoustic}, and random noise attacks to each create 240 adversarial clips of 30 seconds (that 100\% bypassed the surrogate detector), and then uploaded them to a private YouTube channel to test YouTube's copyright detection.

\noindent\textbf{Pairing Attack Success Rates with qDev Values:}
It is clear that we can always get a 100\% attack success rate by generating a sufficiently large perturbation and adding it to the original music, which can, unfortunately, produce extremely noisy sound. Hence, it is necessary to pair the attack success rate with perceptual quality.


\begin{figure}[!t]
    \centering
    \includegraphics[width=0.46\textwidth]{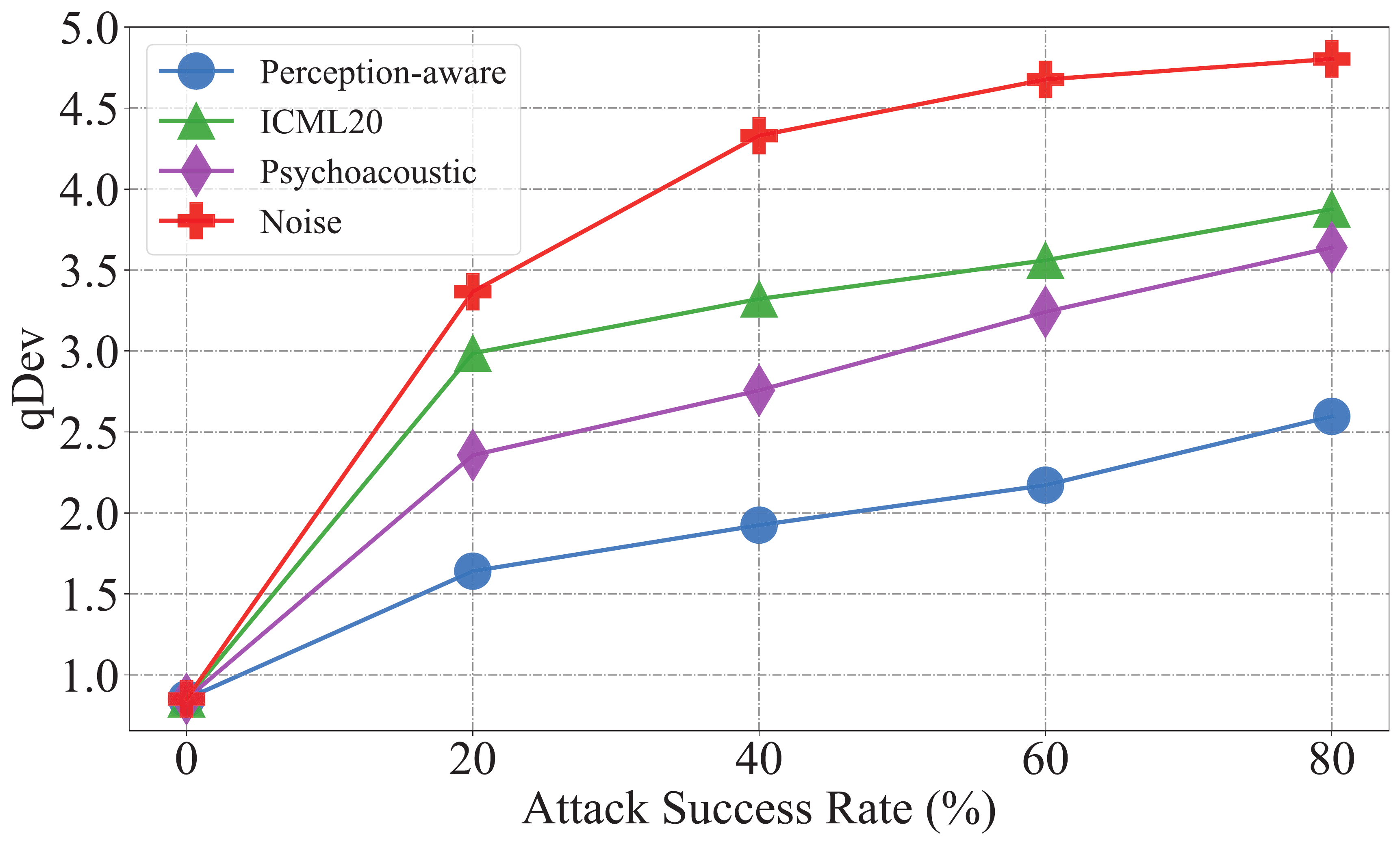}
    \vspace{-0.4cm}
    \caption{\textcolor{black}{Attack success rates pairing with qDev.}}
    \label{Fig:ASR_3method}
\end{figure}


To this end, we focus on comparing the average qDev values of adversarial music clips created by perception-aware, ICML20, \textcolor{black}{psychoacoustic}, and random noise attacks under the same attack success rates against YouTube. Fig.~\ref{Fig:ASR_3method} shows the comparison results. As shown in Fig.~\ref{Fig:ASR_3method}, higher attack success rates come with lower music perceptual quality in general. The qDev values of the perception-aware attack are always better than ICML20, \textcolor{black}{psychoacoustic} and random noise attacks for the same attack success rate. In particular, its qDev increases from 1.64 (good quality with quiet noise) to 2.53 (noticeable with slight noise) when the attack success rate goes from 20\% to 80\%; in contrast, the ICML20 attack has the qDev value increasing from 2.70 (noticeable with slight noise) to nearly 4 (very noisy). \textcolor{black}{The qDev of the psychoacoustic attack ranges from 2.30 to 3.60, exhibiting better performance than ICML20 via its strategy to limit the energy within certain frequencies to suppress human attention.} The random noise attack has the highest qDev value almost reaching 5 when the attack success rate is 80\%. Overall, Fig.~\ref{Fig:ASR_3method} offers very intuitive comparisons and demonstrates that the perception-aware attack is able to create more effective attacks against YouTube with better music quality.

\noindent\textbf{Impact of Number of Instrumental Tracks:}
In our experiments for the perception-aware attack, the number of instruments used to generate the perturbation was set to be $K\!=\!7$. It means that \eqref{Eq:deviation2} always tries to find 7 weights assigned to 7 instrumental tracks. We can reduce the computational complexity by restricting the number of instrumental tracks. The less the number, the less the computational complexity \eqref{Eq:deviation2} incurs. We conducted experiments to evaluate the impact of this number. Specifically, we still used 7 instruments but only choose 1, 3, or 5 out of 7 to form the instrumental track(s) as the perturbations to create the adversarial music clips. Under approximately the same attack success rates against YouTube, we show the average qDev values of 160 adversarial music clips for each various instrument selection method in Table~\ref{tab:remixing_evaluation2}.

\begin{table}[h]
  \caption{Attack success rates and qDev values for different numbers of instruments. }
  \label{tab:remixing_evaluation2}
  \vspace{\TableCaptionSpacing}
  \begin{adjustbox}{center,max width=\linewidth}
    \begin{tabular}{lcccccccc}
      \toprule
      \bf Number of instruments: 
      & 1
      & 3
      & 5
      & 7\\
      \midrule
      \bf Success rate:      & 78.13\%  & 80.00\%     & 79.38\%    & 80.63\%      \\
      \bf qDev:     & 2.8901   & 2.7256  & 2.6713  &2.5902\\
      \bottomrule
    \end{tabular}
  \end{adjustbox}
\end{table}


We find in Table~\ref{tab:remixing_evaluation} that the qDev value gradually decreases from 2.8901 to 2.5902 when we choose 1 to 7 out of 7 instruments to create the perturbations. This is expected as the objective of \eqref{Eq:deviation2} is to minimize qDev and more instrument selections lead to a lower qDev value. One interesting observation is that choosing fewer instruments does not quite affect the attack success rate against YouTube. However, using only one instrument creates a quite loud music signal played by the instrument that is more identifiable to humans. Adding more instruments and distributing weights among them help suppress one single loud perturbation signal and makes the overall perturbation less identifiable.

{\hlb
\subsection{Manipulating Other Music Features}\label{subsec:manipulation_method}
As we have discussed in Section~\ref{Sec:Attack}, our perception-aware attack mainly focuses on generating perturbation via revising the timbre feature of music, which should be more effective than changing pitch or rhythm. But it is certainly feasible to focus on manipulating pitch or rhythm to generate perturbation. Here, we evaluate the perturbations created by manipulating each of the pitch, rhythm, and timbre features.

There is still an open space to manipulate pitch or rhythm with potential optimizations. We adopt a randomized strategy to compare the three manipulations. In particular, we create pitch-based perturbations with a random energy via shifting music notes in its spectrogram by a random frequency, rhythm-based perturbations with a random energy via speeding up and slowing down the tempo of music notes at a random rate, and timbre-based perturbations with a random energy by randomly choosing one instrumental track playing the same music notes. Because of distinct natures in different generations, we must compare them under the same standard. We choose the qDev as the standard, and compare the attack success rates of randomly generated perturbations that always have the same qDev value.

\begin{table}[h]
\caption{\textcolor{black}{Attack success rates with different manipulations.} }
  \label{tab:3method_manipulation}
\vspace{\TableCaptionSpacing}
\begin{adjustbox}{center,max width=\linewidth}
\begin{tabular}{l|cccc}
\toprule
{\bf qDev value:}  & 1.5 & 2.5 & 3.5 & 4.5 \\ \hline
{\bf Pitch:}  & 9.38\%       &20.31\%      & 29.69\%     & 39.06\%       \\
{\bf Rhythm:}  & 7.81\%       & 15.63\%      & 28.13\%     & 54.68\%       \\
{\bf Timbre:}  & 14.06\%      & 31.25\%     & 48.44\%     & 70.31\%      \\ \bottomrule
\end{tabular}
\end{adjustbox}
\end{table}


Table~\ref{tab:3method_manipulation} shows the attack success rates of adversarial music clips created by the three randomized manipulation methods against YouTube under the different qDev values (64 clips for each manipulation method under each qDev level). We can see that the timbre manipulation always achieves higher success rates than pitch and rhythm manipulations in randomized generations. 

Note that it is possible to further optimize the pitch or rhythm manipulation, or even combine all features to formulate a joint framework to minimize the qDev. However, involving them together may incur more search complexity. A balanced manipulation method among multiple features is also worth further studies.
}

\subsection{Discussions}\label{sec:discussion}
Though the perception-aware attack produce better-quality perturbations, we can still notice deviations (some are minor and others more noticeable) from the perturbed music. One may further improve the attack as discussed below.


\noindent\textbf{Subtlety in small qDev difference:} The metric of qDev based on current data regression of human ratings is not sufficiently sensitive to a small value difference. For example, a qDev value decrease from 4 to 1 should indicate an evident music perceptual quality improvement; however, a decrease from 2.1 to 2.0 may well fall into the error range of subjective judgements and is not fully correlated with music quality improvement. This may indicate that within this subtle qDev range, there might exist other improvements to make the perturbation sound more natural and attached to the original music. For example, some instruments (e.g., trumpet during our observations) can produce audio characteristics more identifiable to humans than some others, making its track evidently comparable to the foreground tracks (e.g., the main vocal track) in the original music. It may be necessary for \eqref{Eq:deviation2} to select such an instrument to beat the classification via creating more timbre variations and minimize the qDev. There may exist other benchmarks in this case to further differentiate the selection of instruments as a small qDev difference may no longer help the selection.

\noindent\textbf{Transition in dynamic clipping:} Dynamic clipping segments a music signal into multiple clips and finds the optimal additional instruments for each clip. When the instrument sets for adjacent clips are chosen in a distinct way, human participants may be sharp enough to notice an instrumental transition. Smoothing this transition may result in a better experience; but the smoothing still needs to take suppressing audio fingerprints into consideration.

{\hlb
\noindent\textbf{Robustness and bias of the regression model:}
Our human study and evaluations show that Random Forest achieves the minimum MSEs in both G1 and G2 compared with other models. Based on the Random Forest-regressed qDev, the perception-aware attack achieves better performance than the ICML20 and psychoacoustic attacks. As the human participants in our study are all college students with ages 20-35 and non-music experts, we acknowledge that our perceptual evaluations do not reflect music experts' judgements but show the opinions of general young populations. Extending the perceptual evaluations to other groups (e.g., elder populations and music experts) will help create more accurate and robust prediction.

\noindent\textbf{Genearalizability to speech:} The research in this paper focuses on the music domain. Our general human-in-the-loop methodology can be extended to the speech domain. As there are technical differences between fingerprinting music and recognizing speech, we expect this leads to non-trivial efforts to rebuild the qDev model based on human perception of speech difference, perform sensitivity analysis for speech features, and then shrink the search space by considering qDev-friendly acoustic signals (e.g., from a set of synthetic speech phonemes) to minimize the non-differentiable qDev, which is worth further studies and evaluations.
}

\noindent\textbf{Vulnerability disclosure:}
The perception-aware attack does not cause an immediate operational impact, such as denial of service. Following the practice of responsible disclosure, we reported the issue of music copyright detection to Google. Google initially classified the case as an abuse risk. During the commmunicaiton, Google mentioned that a copyright content will be taken down from YouTube when the copyright owner makes a request. Google eventually made the decision not to track it as a security bug.

\section{Discussions on Defense Strategies}\label{Sec:Defense}
In this section, we discuss potential defense strategies.

{\vspace{0.1cm}\noindent\bf Existing audio defense:} Audio pre-processing is a potential method to reduce the effectiveness of adversarial examples, as the small perturbation could be mitigated during the audio squeezing \cite{yuan2018commandersong,chen2020devil,chen2019real,zheng2021black} and audio compression \cite{li2020advpulse,das2018adagio}. These defense methods are unlikely effective against the perception-aware attack as  squeezing/compression does not quite change the spectrogram feature (e.g., the high energy harmonics will not be revised during the processing). On the other hand, these defense methods may not be desirable in some scenarios. For example, YouTube does not downgrade the music quality via squeezing and compression. 

{\vspace{0.1cm}\noindent\bf {Improving audio fingerprinting:}}  The advantage of audio fingerprinting is its computational efficiency \cite{haitsma2002highly,gomez2002mixed,wang2003industrial}. Exiting research \cite{wang2003industrial,fenet2011scalable, sonnleitner2015robust} focused mostly on extracting spectrogram features in a robust way for fingerprinting based detection. Although these fingerprints can be made robust to noise and pitch-shifting \cite{fenet2011scalable}, the perception-aware attack creates additional harmonics and spectrogram features that can be extracted as fingerprints and fool the detection. We can potentially improve audio fingerprinting against the perception-aware attack by adding the pitch and rhythm features as other types of fingerprints. This, however, will incur substantially more costs because estimating pitch and rhythm incurs complicated maximum-likelihood estimation \cite{duan2010multiple} than spectrogram based fingerprinting. There is a need to achieve a balanced tradeoff between detection accuracy and computational complexity. 

{\vspace{0.1cm}\noindent\bf Defense in machine learning:}
Another possible way to defend against the perception-aware attack is to leverage existing defense strategies from the machine learning community. In particular, adversarial training \cite{goodfellow2014explaining, madry2017towards, balaji2019instance, cai2018curriculum, shafahi2019adversarial, tramer2017ensemble, wong2020fast} and certified defense \cite{balunovic2019adversarial, wong2018provable, mirman2018differentiable,katz2017reluplex} are popular among the methods to provide more robustness against adversarial attacks. Adversarial training primarily focuses on making the model robust to the adversaries via solving a min-max optimization problem that finds the model parameters to minimize the cost results from strong adversary examples. Given a bounded $L_p$ ball, the re-trained model becomes more robust against the adversarial attacks. However, the perception-aware attacker uses qDev instead of $L_p$ norm to craft adversarial examples. This creates a model mismatch \cite{sharma2017attacking} and can make the re-trained model ill-suited. A potential way to solve the issue is to use qDev to guide the adversarial training. However, computing qDev is a non-differentiable process. Initial efforts can be focused on finding a differentiable function to approximate qDev to efficiently finish the adversarial training. Certified defense is to find an upper bound of the adversarial loss which guarantees the robustness to any attack in the same threat model. Existing work \cite{balunovic2019adversarial} can provide a provable defense to the neural networks via convex layerwise adversarial training. To use certified defense against the perception-aware attack, we need to find a differential upper bound to characterize the adversarial loss based on the qDev modeling, which, similar to using adversarial training, involves non-trivial research efforts.



\section{Related Work}\label{Sec:RelatedWork}

{\vspace{0.0cm}\noindent\bf Adversarial audio attacks:}
Most adversarial attacks \cite{carlini2018audio,kreuk2018fooling,chen2019real,li2020advpulse,zheng2021black} control the energy of the perturbation within a bounded $L_p$ ball such that a created adversarial audio example resembles the original signal in its waveform format. In this paper, we show that limiting the waveform change is not fully related to human-perceived change. Instead of using the $L_p$ norm, we propose to use qDev based on the comprehensive human study to create adversarial signals with better quality.  
There are also a few recent studies \cite{zhang2017dolphinattack, carlini2016hidden,abdullah2019practical, yuan2018commandersong,chen2020devil} focusing on creating inaudible or stealthy signals as attacks. 
These studies generally use various strategies to effectively hide the presence of the attack. The perception-aware attack adopts a different strategy that creates perturbation signals to minimize the human-perceived deviation. The ICML20 method \cite{saadatpanah2020adversarial} focused on creating a neural network based black-box attack against copyright detectors. It proposed a mathematical attempt that enforces the perturbation to be similar to a signal of certain frequencies to make it more natural based on $L_p$ norm. {\hlb Several studies on speech recognition attacks \cite{schonherr2018adversarial,qin2019imperceptible,lireal} also presented psychoacoustic hiding methods to embed low energy perturbations near the frequency of a louder signal to improve the perceptual quality. We have adapted the approach in \cite{qin2019imperceptible} to create music perturbations. Compared with ICML20 and psychoacoustic attacks, the perception-aware attack integrates the proposed qDev into its formulation, and creates effective adversarial music while achieving better perceptual quality in our evaluations}.

{\vspace{0.1cm}\noindent\bf Human evaluation of audio quality:}
Human perception studies \cite{carlini2016hidden,yuan2018commandersong,chen2019real,chen2020devil,zheng2021black} have been adopted to evaluate the stealthiness of adversarial audio examples as the SNR metric may not be appropriate to well reflect the human perception \cite{chen2020devil,zheng2021black}. Exiting work \cite{carlini2016hidden,yuan2018commandersong,chen2019real,chen2020devil,zheng2021black} designed human perception studies from different perspectives and evaluated the attack performance based on the results of human study. For instance, \cite{wenger2021hello} conducted a comprehensive human study to evaluate the synthetic speech quality to reveal the impact of deep-learning based speech synthesis to human. These studies focused on analyzing the results of the human evaluation, rather than integrating human factors into the designs. There are few studies \cite{tsai2011automatic,gupta2017perceptual} focusing on defining human-involved metrics for singing scoring systems. The systems were designed to generate an absolute score to indicate the singing performance given the recording of a human's singing via linear weighting \cite{tsai2011automatic} or non-linear neural network \cite{gupta2017perceptual} on audio features. By contrast, our strategy focuses on modeling the human-perceived deviation between original and perturbed music signals, compares different regression models, and analyzes how each audio feature affects the overall human perception of music deviation.

\section{Conclusion}\label{Sec:Conclusion}
In this paper, we conducted a human study to reverse-engineer the human perception of music deviation via regression analysis. Based on the analysis, we proposed the perception-aware attack framework to create adversarial music that can mislead a music classifier while preserving the perceptual quality. Experimental results have shown that the perception-aware attack is effective and achieves better music perceptual quality compared to prior work. Our work demonstrates that perceptual quality of adversarial attacks can be significantly improved by integrating human factors into the adversarial audio attack design process.

\bibliographystyle{plain}
\bibliography{arxiv_version}

\begin{thebibliography}{10}

\bibitem{test_amazon}
{Amazon Alexa}.
\newblock \url{https://developer.amazon.com/en-US/alexa}, 2022.
\newblock Accessed: 2022-01-07.

\bibitem{test_Google}
{Google Assistant}.
\newblock \url{https://assistant.google.com/}, 2022.
\newblock Accessed: 2022-01-07.

\bibitem{abdullah2019practical}
Hadi Abdullah, Washington Garcia, Christian Peeters, Patrick Traynor, Kevin~RB
  Butler, and Joseph Wilson.
\newblock Practical hidden voice attacks against speech and speaker recognition
  systems.
\newblock {\em In Proc. of NDSS}, 2019.

\bibitem{abdullah2019hear}
Hadi Abdullah, Muhammad~Sajidur Rahman, Washington Garcia, Logan Blue, Kevin
  Warren, Anurag~Swarnim Yadav, Tom Shrimpton, and Patrick Traynor.
\newblock Hear" no evil", see" kenansville": Efficient and transferable
  black-box attacks on speech recognition and voice identification systems.
\newblock {\em In Proc. of IEEE S\&P}, 2021.

\bibitem{allamanche2001audioid}
Eric Allamanche.
\newblock Audioid: Towards content-based identification of audio material.
\newblock In {\em Proc. of AES}, 2001.

\bibitem{bailis2005mortality}
Robert Bailis, Majid Ezzati, and Daniel~M Kammen.
\newblock Mortality and greenhouse gas impacts of biomass and petroleum energy
  futures in africa.
\newblock {\em In Proc. of Science}, 2005.

\bibitem{balaji2019instance}
Yogesh Balaji, Tom Goldstein, and Judy Hoffman.
\newblock Instance adaptive adversarial training: Improved accuracy tradeoffs
  in neural nets.
\newblock {\em arXiv preprint arXiv:1910.08051}, 2019.

\bibitem{balunovic2019adversarial}
Mislav Balunovic and Martin Vechev.
\newblock Adversarial training and provable defenses: Bridging the gap.
\newblock In {\em Proc. of ICLR}, 2019.

\bibitem{boney1996digital}
Laurence Boney, Ahmed~H Tewfik, and Khaled~N Hamdy.
\newblock Digital watermarks for audio signals.
\newblock In {\em Proc. of ICMCS}, 1996.

\bibitem{cai2018curriculum}
Qi-Zhi Cai, Min Du, Chang Liu, and Dawn Song.
\newblock Curriculum adversarial training.
\newblock {\em In Proc. of IJCAI}, 2018.

\bibitem{campbell2008photosynthetic}
J~Elliott Campbell, Gregory~R Carmichael, T~Chai, M~Mena-Carrasco, Y~Tang,
  DR~Blake, NJ~Blake, Stephanie~A Vay, G~James Collatz, I~Baker, et~al.
\newblock Photosynthetic control of atmospheric carbonyl sulfide during the
  growing season.
\newblock {\em In Proc. of Science}, 2008.

\bibitem{cano2005review}
Pedro Cano, Eloi Batlle, Ton Kalker, and Jaap Haitsma.
\newblock A review of audio fingerprinting.
\newblock {\em Journal of VLSI signal processing systems for signal, image and
  video technology}, 41(3):271--284, 2005.

\bibitem{cano2002robust}
Pedro Cano, Eloi Batlle, Harald Mayer, and Helmut Neuschmied.
\newblock Robust sound modeling for song detection in broadcast audio.
\newblock {\em In Proc. AES 112th Int. Conv}, 2002.

\bibitem{carlini2016hidden}
Nicholas Carlini, Pratyush Mishra, Tavish Vaidya, Yuankai Zhang, Micah Sherr,
  Clay Shields, David Wagner, and Wenchao Zhou.
\newblock Hidden voice commands.
\newblock In {\em Proc. of USENIX Security}, 2016.

\bibitem{carlini2017towards}
Nicholas Carlini and David Wagner.
\newblock Towards evaluating the robustness of neural networks.
\newblock In {\em Proc. of IEEE S\&P}, 2017.

\bibitem{carlini2018audio}
Nicholas Carlini and David Wagner.
\newblock Audio adversarial examples: Targeted attacks on speech-to-text.
\newblock In {\em Proc. of SPW}, 2018.

\bibitem{casey2008content}
Michael~A Casey, Remco Veltkamp, Masataka Goto, Marc Leman, Christophe Rhodes,
  and Malcolm Slaney.
\newblock Content-based music information retrieval: Current directions and
  future challenges.
\newblock {\em In Proc. of IEEE}, 2008.

\bibitem{chen2019real}
Guangke Chen, Sen Chen, Lingling Fan, Xiaoning Du, Zhe Zhao, Fu~Song, and Yang
  Liu.
\newblock Who is real bob? adversarial attacks on speaker recognition systems.
\newblock {\em In Proc. of IEEE S\&P}, 2021.

\bibitem{chen2020devil}
Yuxuan Chen, Xuejing Yuan, Jiangshan Zhang, Yue Zhao, Shengzhi Zhang, Kai Chen,
  and XiaoFeng Wang.
\newblock Devil’s whisper: A general approach for physical adversarial
  attacks against commercial black-box speech recognition devices.
\newblock In {\em Proc. of USENIX Security}, 2020.

\bibitem{cox2002digital}
Ingemar~J Cox, Matthew~L Miller, Jeffrey~Adam Bloom, and Chris Honsinger.
\newblock {\em Digital watermarking}, volume~53.
\newblock Springer, 2002.

\bibitem{daniel1987spearman}
Wayne~W Daniel.
\newblock The spearman rank correlation coefficient.
\newblock {\em In Proc. of Biostatistics: A Foundation for Analysis in the
  Health Sciences}, 1987.

\bibitem{das2018adagio}
Nilaksh Das, Madhuri Shanbhogue, Shang-Tse Chen, Li~Chen, Michael~E Kounavis,
  and Duen~Horng Chau.
\newblock Adagio: Interactive experimentation with adversarial attack and
  defense for audio.
\newblock In {\em Joint European Conference on Machine Learning and Knowledge
  Discovery in Databases}, pages 677--681. Springer, 2018.

\bibitem{davis1980comparison}
Steven Davis and Paul Mermelstein.
\newblock Comparison of parametric representations for monosyllabic word
  recognition in continuously spoken sentences.
\newblock {\em In Proc. of IEEE TASSP}, 1980.

\bibitem{de2012enhancing}
Franz De~Leon and Kirk Martinez.
\newblock Enhancing timbre model using mfcc and its time derivatives for music
  similarity estimation.
\newblock In {\em Proc. of EUSIPCO}, 2012.

\bibitem{duan2010multiple}
Zhiyao Duan, Bryan Pardo, and Changshui Zhang.
\newblock Multiple fundamental frequency estimation by modeling spectral peaks
  and non-peak regions.
\newblock {\em IEEE Transactions on Audio, Speech, and Language Processing},
  18(8):2121--2133, 2010.

\bibitem{fenet2011scalable}
S{\'e}bastien Fenet, Ga{\"e}l Richard, Yves Grenier, et~al.
\newblock A scalable audio fingerprint method with robustness to
  pitch-shifting.
\newblock In {\em Proc. of ISMIR}, pages 121--126, 2011.

\bibitem{godsill2002bayesian}
Simon Godsill and Manuel Davy.
\newblock Bayesian harmonic models for musical pitch estimation and analysis.
\newblock In {\em IEEE International Conference on Acoustics, Speech, and
  Signal Processing}, volume~2, pages II--1769. IEEE, 2002.

\bibitem{godsill2003bayesian}
SIMON~J Godsill and M~Davy.
\newblock Bayesian harmonic models for musical signal analysis.
\newblock {\em In Proc. of Bayesian Statistics}, 7:105--124, 2003.

\bibitem{gomez2002mixed}
Emilia Gomez, Pedro Cano, L~Gomes, Eloi Batlle, and Madeleine Bonnet.
\newblock Mixed watermarking-fingerprinting approach for integrity verification
  of audio recordings.
\newblock In {\em Proc. of $~$ITelCon}, 2002.

\bibitem{goodfellow2014explaining}
Ian~J Goodfellow, Jonathon Shlens, and Christian Szegedy.
\newblock Explaining and harnessing adversarial examples.
\newblock {\em arXiv preprint arXiv:1412.6572}, 2014.

\bibitem{gupta2017perceptual}
Chitralekha Gupta, Haizhou Li, and Ye~Wang.
\newblock Perceptual evaluation of singing quality.
\newblock In {\em Proc. of APSIPA ASC}, pages 577--586, 2017.

\bibitem{gupta2018technical}
Chitralekha Gupta, Haizhou Li, and Ye~Wang.
\newblock A technical framework for automatic perceptual evaluation of singing
  quality.
\newblock {\em In Proc. of APSIPA Transactions on Signal and Information
  Processing}, 7, 2018.

\bibitem{haitsma2002highly}
Jaap Haitsma and Ton Kalker.
\newblock A highly robust audio fingerprinting system.
\newblock In {\em Proc. of Ismir}, volume 2002, pages 107--115, 2002.

\bibitem{haitsma2001robust}
Jaap Haitsma, Ton Kalker, and Job Oostveen.
\newblock Robust audio hashing for content identification.
\newblock In {\em Proc. of CBMIW}, 2001.

\bibitem{hartmann2004signals}
William~M Hartmann.
\newblock {\em Signals, sound, and sensation}.
\newblock In Proc. of Springer Science \& Business Media, 2004.

\bibitem{katz2017reluplex}
Guy Katz, Clark Barrett, David~L Dill, Kyle Julian, and Mykel~J Kochenderfer.
\newblock Reluplex: An efficient smt solver for verifying deep neural networks.
\newblock In {\em International Conference on Computer Aided Verification},
  pages 97--117. Springer, 2017.

\bibitem{kereliuk2007indirect}
Corey Kereliuk, Bertrand Scherrer, Vincent Verfaille, Philippe Depalle, and
  Marcelo~M Wanderley.
\newblock Indirect acquisition of fingerings of harmonic notes on the flute.
\newblock In {\em Proc. of ICMC}, 2007.

\bibitem{kreuk2018fooling}
Felix Kreuk, Yossi Adi, Moustapha Cisse, and Joseph Keshet.
\newblock Fooling end-to-end speaker verification with adversarial examples.
\newblock In {\em Proc. of ICASSP}, pages 1962--1966. IEEE, 2018.

\bibitem{kurakin2016adversarial}
Alexey Kurakin, Ian Goodfellow, Samy Bengio, et~al.
\newblock Adversarial examples in the physical world, 2016.

\bibitem{law2012assessing}
Lily~NC Law and Marcel Zentner.
\newblock Assessing musical abilities objectively: Construction and validation
  of the profile of music perception skills.
\newblock {\em PloS one}, 7(12):e52508, 2012.

\bibitem{lireal}
Juncheng~B Li, Shuhui Qu, Xinjian Li, Zico Kolter, and Florian Metze.
\newblock Real world audio adversary against wake-word detection systems.
\newblock {\em In Proc. of NIPS}.

\bibitem{li2020advpulse}
Zhuohang Li, Yi~Wu, Jian Liu, Yingying Chen, and Bo~Yuan.
\newblock Advpulse: Universal, synchronization-free, and targeted audio
  adversarial attacks via subsecond perturbations.
\newblock In {\em Proc. of ACM CCS}, pages 1121--1134, 2020.

\bibitem{loeffler2006instrument}
Dominik~B Loeffler.
\newblock {\em Instrument timbres and pitch estimation in polyphonic music}.
\newblock PhD thesis, Georgia Institute of Technology, 2006.

\bibitem{madry2017towards}
Aleksander Madry, Aleksandar Makelov, Ludwig Schmidt, Dimitris Tsipras, and
  Adrian Vladu.
\newblock Towards deep learning models resistant to adversarial attacks.
\newblock {\em In Proc. of ICML Work Shop}, 2017.

\bibitem{mirman2018differentiable}
Matthew Mirman, Timon Gehr, and Martin Vechev.
\newblock Differentiable abstract interpretation for provably robust neural
  networks.
\newblock In {\em Proc. of ICML}, pages 3578--3586. PMLR, 2018.

\bibitem{molina2013fundamental}
Emilio Molina, Isabel Barbancho, Emilia G{\'o}mez, Ana~Maria Barbancho, and
  Lorenzo~J Tard{\'o}n.
\newblock Fundamental frequency alignment vs. note-based melodic similarity for
  singing voice assessment.
\newblock In {\em Proc. of International Conference on Acoustics, Speech and
  Signal Processing(ICASSP)}, pages 744--748. IEEE, 2013.

\bibitem{moorer1977signal}
James~Anderson Moorer.
\newblock Signal processing aspects of computer music: A survey.
\newblock {\em In Proc. of the IEEE}, 65(8):1108--1137, 1977.

\bibitem{muller2011signal}
Meinard Muller, Daniel~PW Ellis, Anssi Klapuri, and Ga{\"e}l Richard.
\newblock Signal processing for music analysis.
\newblock {\em IEEE Journal of selected topics in signal processing},
  5(6):1088--1110, 2011.

\bibitem{murphy2004quantification}
James~M Murphy, David~MH Sexton, David~N Barnett, Gareth~S Jones, Mark~J Webb,
  Matthew Collins, and David~A Stainforth.
\newblock Quantification of modelling uncertainties in a large ensemble of
  climate change simulations.
\newblock {\em In Proc. of Nature}, 2004.

\bibitem{neuschmied2001content}
Helmut Neuschmied, Harald Mayer, and Eloi Batlle.
\newblock Content-based identification of audio titles on the internet.
\newblock In {\em Proc. of WEDELMUSIC}, 2001.

\bibitem{pardo2006finding}
Bryan Pardo.
\newblock Finding structure in audio for music information retrieval.
\newblock {\em IEEE Signal Processing Magazine}, 23(3):126--132, 2006.

\bibitem{platel1997structural}
Herv{\'e} Platel, Cathy Price, Jean-Claude Baron, Richard Wise, Jany Lambert,
  Richard~S Frackowiak, Bernard Lechevalier, and Francis Eustache.
\newblock The structural components of music perception. a functional
  anatomical study.
\newblock {\em Brain: a journal of neurology}, 120(2):229--243, 1997.

\bibitem{qin2019imperceptible}
Yao Qin, Nicholas Carlini, Garrison Cottrell, Ian Goodfellow, and Colin Raffel.
\newblock Imperceptible, robust, and targeted adversarial examples for
  automatic speech recognition.
\newblock In {\em Proc. of ICML}, pages 5231--5240. PMLR, 2019.

\bibitem{ratanamahatana2004making}
Chotirat~Ann Ratanamahatana and Eamonn Keogh.
\newblock Making time-series classification more accurate using learned
  constraints.
\newblock In {\em Proceedings of the 2004 SIAM international conference on data
  mining}, pages 11--22. SIAM, 2004.

\bibitem{risset1999exploration}
Jean-Claude Risset and David~L Wessel.
\newblock Exploration of timbre by analysis and synthesis.
\newblock In {\em The psychology of music}, pages 113--169. Elsevier, 1999.

\bibitem{rix2001perceptual}
Antony~W Rix, John~G Beerends, Michael~P Hollier, and Andries~P Hekstra.
\newblock Perceptual evaluation of speech quality (pesq)-a new method for
  speech quality assessment of telephone networks and codecs.
\newblock In {\em Proc. of IEEE International Conference on Acoustics, Speech,
  and Signal Processing(ICASSP)}, volume~2, pages 749--752. IEEE, 2001.

\bibitem{saadatpanah2020adversarial}
Parsa Saadatpanah, Ali Shafahi, and Tom Goldstein.
\newblock Adversarial attacks on copyright detection systems.
\newblock In {\em Proc. of ICML}, pages 8307--8315. PMLR, 2020.

\bibitem{sakoe1978dynamic}
Hiroaki Sakoe and Seibi Chiba.
\newblock Dynamic programming algorithm optimization for spoken word
  recognition.
\newblock {\em IEEE transactions on acoustics, speech, and signal processing},
  26(1):43--49, 1978.

\bibitem{salvador2007toward}
Stan Salvador and Philip Chan.
\newblock Toward accurate dynamic time warping in linear time and space.
\newblock {\em In Proc. of Intelligent Data Analysis}, 11(5):561--580, 2007.

\bibitem{schonherr2018adversarial}
Lea Sch{\"o}nherr, Katharina Kohls, Steffen Zeiler, Thorsten Holz, and Dorothea
  Kolossa.
\newblock Adversarial attacks against automatic speech recognition systems via
  psychoacoustic hiding.
\newblock {\em In Proc. of NDSS}, 2019.

\bibitem{sedgwick2014spearman}
Philip Sedgwick.
\newblock Spearman’s rank correlation coefficient.
\newblock {\em In Proc, of Bmj}, 349, 2014.

\bibitem{shafahi2019adversarial}
Ali Shafahi, Mahyar Najibi, Amin Ghiasi, Zheng Xu, John Dickerson, Christoph
  Studer, Larry~S Davis, Gavin Taylor, and Tom Goldstein.
\newblock Adversarial training for free!
\newblock {\em In Proc. of NIPS}, 2019.

\bibitem{sharma2017attacking}
Yash Sharma and Pin-Yu Chen.
\newblock Attacking the madry defense model with $ l\_1 $-based adversarial
  examples.
\newblock {\em In Proc. of ICLR Work Shop}, 2018.

\bibitem{sonnleitner2015robust}
Reinhard Sonnleitner and Gerhard Widmer.
\newblock Robust quad-based audio fingerprinting.
\newblock {\em IEEE/ACM Transactions on Audio, Speech, and Language
  Processing}, 24(3):409--421, 2015.

\bibitem{szegedy2013intriguing}
Christian Szegedy, Wojciech Zaremba, Ilya Sutskever, Joan Bruna, Dumitru Erhan,
  Ian Goodfellow, and Rob Fergus.
\newblock Intriguing properties of neural networks.
\newblock {\em arXiv preprint arXiv:1312.6199}, 2013.

\bibitem{thiede2000peaq}
Thilo Thiede, William~C Treurniet, Roland Bitto, Christian Schmidmer, Thomas
  Sporer, John~G Beerends, and Catherine Colomes.
\newblock Peaq-the itu standard for objective measurement of perceived audio
  quality.
\newblock {\em Journal of the Audio Engineering Society}, 48(1/2):3--29, 2000.

\bibitem{tramer2017ensemble}
Florian Tram{\`e}r, Alexey Kurakin, Nicolas Papernot, Ian Goodfellow, Dan
  Boneh, and Patrick McDaniel.
\newblock Ensemble adversarial training: Attacks and defenses.
\newblock {\em In Proc. of ICLR}, 2018.

\bibitem{tsai2011automatic}
Wei-Ho Tsai and Hsin-Chieh Lee.
\newblock Automatic evaluation of karaoke singing based on pitch, volume, and
  rhythm features.
\newblock {\em IEEE Transactions on Audio, Speech, and Language Processing},
  20(4):1233--1243, 2011.

\bibitem{uhlich2017improving}
Stefan Uhlich, Marcello Porcu, Franck Giron, Michael Enenkl, Thomas Kemp, Naoya
  Takahashi, and Yuki Mitsufuji.
\newblock Improving music source separation based on deep neural networks
  through data augmentation and network blending.
\newblock In {\em Proc. of ICASSP}, pages 261--265. IEEE, 2017.

\bibitem{valentini2016investigating}
Cassia Valentini-Botinhao, Xin Wang, Shinji Takaki, and Junichi Yamagishi.
\newblock Investigating rnn-based speech enhancement methods for noise-robust
  text-to-speech.
\newblock In {\em SSW}, pages 146--152, 2016.

\bibitem{vaseghi2008advanced}
Saeed~V Vaseghi.
\newblock {\em Advanced digital signal processing and noise reduction}.
\newblock John Wiley \& Sons, 2008.

\bibitem{walmsley1998multidimensional}
Paul~J Walmsley, Simon~J Godsill, and Peter~JW Rayner.
\newblock Multidimensional optimisation of harmonic signals.
\newblock In {\em 9th European Signal Processing Conference (EUSIPCO 1998)},
  pages 1--4. IEEE, 1998.

\bibitem{wang2003industrial}
Avery Wang et~al.
\newblock An industrial strength audio search algorithm.
\newblock In {\em Proc. of Ismir}, volume 2003, pages 7--13. Washington, DC,
  2003.

\bibitem{wenger2021hello}
Emily Wenger, Max Bronckers, Christian Cianfarani, Jenna Cryan, Angela Sha,
  Haitao Zheng, and Ben~Y Zhao.
\newblock " hello, it's me": Deep learning-based speech synthesis attacks in
  the real world.
\newblock In {\em Proc. of ACM CCS}, pages 235--251, 2021.

\bibitem{wessel1979timbre}
David~L Wessel.
\newblock Timbre space as a musical control structure.
\newblock {\em Computer music journal}, pages 45--52, 1979.

\bibitem{wong2018provable}
Eric Wong and Zico Kolter.
\newblock Provable defenses against adversarial examples via the convex outer
  adversarial polytope.
\newblock In {\em Proc. of ICML}, pages 5286--5295. PMLR, 2018.

\bibitem{wong2020fast}
Eric Wong, Leslie Rice, and J~Zico Kolter.
\newblock Fast is better than free: Revisiting adversarial training.
\newblock {\em arXiv preprint arXiv:2001.03994}, 2020.

\bibitem{yakura2018robust}
Hiromu Yakura and Jun Sakuma.
\newblock Robust audio adversarial example for a physical attack.
\newblock {\em Proc. of IJCAI}, 2018.

\bibitem{yuan2018commandersong}
Xuejing Yuan, Yuxuan Chen, Yue Zhao, Yunhui Long, Xiaokang Liu, Kai Chen,
  Shengzhi Zhang, Heqing Huang, XiaoFeng Wang, and Carl~A Gunter.
\newblock Commandersong: A systematic approach for practical adversarial voice
  recognition.
\newblock In {\em Proc. of USENIX Security}, 2018.

\bibitem{zhang2017dolphinattack}
Guoming Zhang, Chen Yan, Xiaoyu Ji, Tianchen Zhang, Taimin Zhang, and Wenyuan
  Xu.
\newblock Dolphinattack: Inaudible voice commands.
\newblock In {\em Proc. of ACM CCS}, pages 103--117, 2017.

\bibitem{zheng2021black}
Baolin Zheng, Peipei Jiang, Qian Wang, Qi~Li, Chao Shen, Cong Wang, Yunjie Ge,
  Qingyang Teng, and Shenyi Zhang.
\newblock Black-box adversarial attacks on commercial speech platforms with
  minimal information.
\newblock {\em In Proc. of ACM CCS}, 2021.

\end{thebibliography}

\end{document}